\documentclass[nofootinbib,eqsecnum,tightenlines,superscriptaddress,10pt]{revtex4}

\usepackage{graphicx}
\usepackage{amsmath,amssymb,amsfonts,amsthm,stmaryrd,mathtools,bm}
\usepackage{mathrsfs}
\usepackage{color}
\usepackage{multirow,bigdelim}
\usepackage{hyperref}
\usepackage{dsfont}
\usepackage{booktabs}
\allowdisplaybreaks[0]

\usepackage{pstricks}
\usepackage{tikz}
\usepackage{ulem}

\usepackage{caption}
\usepackage{subcaption}

\def\be#1\ee{\begin{align}#1\end{align}}

\def\ba{\begin{eqnarray}}
\def\ea{\end{eqnarray}}
\def\nn{\nonumber}
\def\q{\quad}
\def\bd{\mathbf{d}}

\newcommand{\BD}[1]{{\leavevmode {#1}}}

\begin{document}

\title{Lorentzian quantum cosmology from effective spin foams}

\author{Bianca Dittrich}
\affiliation{Perimeter Institute, 31 Caroline Street North, Waterloo, ON, N2L 2Y5, Canada}
\author{Jos\'e Padua-Arg\"uelles}
\affiliation{Perimeter Institute, 31 Caroline Street North, Waterloo, ON, N2L 2Y5, Canada}
\affiliation{Department of Physics and  Astronomy, University of Waterloo, 200 University Avenue West, Waterloo, ON, N2L 3G1, Canada}

\begin{abstract}

Effective spin foams  provide the  computationally most efficient spin foam models yet and are therefore ideally suited  for applications e.g. to quantum cosmology.  We provide here the first effective spin foam computations of a finite time evolution step in a Lorentzian quantum de Sitter universe. We will consider a set-up which computes the no-boundary wave function, as well as a set-up describing the transition between two finite scale factors.   
A key property of spin foams is that they implement discrete spectra for the areas. We therefore study the effects that are induced by the discrete spectra.

To perform these computations we had to identify a technique to deal with highly oscillating and slowly converging, or even diverging sums. We illustrate here that high order Shanks transformation work very well and are a promising tool for the evaluation of Lorentzian (gravitational) path integrals and spin foam sums.


\end{abstract}

\maketitle

\section{Introduction}

The real time or Lorentzian path integral is in particular important for quantum gravity \cite{deBoer,CDT,EPRLFK,PerezReview,TurokEtAl,EffSF3,Sato}. Euclidean quantum gravity approaches rely on a  formal Wick rotation, to justify a path integral based on the Euclidean gravity action, over Euclidean metrics. Such Euclidean approaches suffer however from the conformal factor problem \cite{ConfFac}.  Furthermore, the space of Euclidean metrics is quite different from the space of Lorentzian metrics, making the notion of (inverse) Wick rotation in general  ill-defined \cite{CDT}. 

The Lorentzian path integral is however very challenging to compute. The oscillating amplitude prohibits the application of Monte Carlo methods, and a direct integration or summation typically leads to a very slow convergence. One way to deal with this issue is to deform the integration contour into the complex plane, using e.g. Picard-Lefschetz methods \cite{PLTheory,Witten1,RelTime,QCDReview,TurokEtAl,HanLefschetz,Ding2021,ADP}.  

In this work we will consider Lorentzian effective spin foams applied to a cosmological model describing de Sitter space.    \BD{The spin foam approach \cite{PerezReview} provides a path integral for quantum gravity, based on the quantum geometric ingredients of loop quantum gravity \cite{books}. The derivation of a cosmological dynamics from spin foams is arguably much less developed than loop quantum cosmology \cite{LQCreviews}, which studies loop quantum gravity versions of the Wheeler-DeWitt equation. There are however a number of important open questions in loop quantum cosmology, including whether the theory is fully covariant \cite{Bojowald} and how loop quantum cosmology relates to the full theory of loop quantum gravity \cite{Brunnemann:2007du}. Spin foams can serve as an independent approach to quantum cosmology \cite{SFC1} and might help to answer these questions. A key difficulty for spin foams are however the extremely high computational demands \cite{Dona2022} of models such as the EPRL/FK model \cite{EPRLFK}. Effective spin foams, introduced in \cite{EffSF1,EffSF2,EffSF3} require far less computational resources. As this paper will illustrate, effective spin foams allow to probe quantum time evolution in cosmology, which so far has not been achieved in spin foam cosmology using other spin foam models \cite{SFC1,SFC2,SFC3,SFC4,SFC5}.}

In contrast to previous works on spin foam cosmology \cite{SFC1,SFC2,SFC3,SFC4,SFC5}, the spin foam partition function discussed in this work will include an unbounded  sum over a lapse-like bulk variable. \BD{This sum features an oscillating amplitude and is not necessarily absolutely convergent. A key issue which we therefore face here, is how to perform this sum.  The corresponding path integrals for Regge calculus \cite{ADP} or for continuum mini-superspace \cite{TurokEtAl} result in proper integrals, and, as mentioned above, can be computed using a deformation of the integration contour in a complexified configurations space, or more specifically Picard-Lefschetz theory \cite{PLTheory,Witten1}. But these techniques are not applicable to sums \cite{Unpublished}.
}

\BD{Here we will show that one can treat such sums arising from partition functions with oscillating amplitudes with non-linear series transformations} \cite{WenigerReview}, and in particular Wynn's epsilon algorithm, which accelerates the convergence of these sums.  Surprisingly, this technique can also be applied to evaluate expectation values, which involve diverging sums. We identify as reason for the effectiveness of these techniques the simple asymptotic behaviour of the gravitational (Regge) action. This gives hope that these techniques can be applied for more general gravitational and possibly non-gravitational path integrals and sums.

\BD{A key question we will pursue here, is how the replacement of integrals into sums affects the dynamics. These sums arise due to (loop) quantum geometry, which prescribes a discrete spectrum for areas \cite{Smolin,Ashtekar,Conrady}, and it is important to understand how the discreteness of geometric variables such as areas does influence the dynamics. }

\BD{Effective spin foams \cite{EffSF1,EffSF2,EffSF3} do implement the key principles of spin foam construction, in particular a discrete area spectrum.} They offer two advantages over previous spin foam models \cite{PerezReview}, which make them in particular suitable for applications to cosmology. The first advantage is that their numerical evaluations is by several magnitudes faster than for EPRL/FK models \cite{EPRLFK}.  Here we will provide the first evaluation of cosmological spin foams with an internal bulk (area) variable. One such evaluation on a standard laptop can take less than a second. In contrast EPRL models require high performance computational resources and have led to the development of specialized algorithms for their evaluation \cite{Dona2022}. Despite these concentrated efforts the most recent works, which evaluate EPRL spin foam amplitudes for cosmology \cite{SFC4,SFC5} via high performance computation, do not involve a summation over a bulk area variable.\footnote{\cite{SFC4} considers the amplitude for a four-simplex and \cite{SFC5} applies a refining 1-4 Pachner move to each of the five boundary tetrahedra of the four-simplex. This does lead to five bulk tetrahedra but no bulk triangles.} 

Another important advantage of effective spin foams is that they allow sub-simplices of arbitrary signatures, in particular time-like tetrahedra, triangles and edges. In contrast, the standard EPRL/FK models rely in their construction on the so-called time gauge, which forces all tetrahedra (and thus triangles and edges) to be space-like. An extension of the EPRL/FK models which includes more general building blocks does exist \cite{Conrady}, but the development of numerical techniques for these extended formulation lags very much behind the ones for the standard formulation \cite{Dona2022}.  Many cosmological mini-superspace path integrals incorporate an integral over a lapse like variable (which then might be Wick-rotated). Such a lapse like variable is difficult to construct if only space-like tetrahedra are allowed. 

The effective spin foam models do rely in their construction more directly on the Regge action. The models described here will automatically incorporate the correct (semi-)classical limit. This point has been a main concern for spin foam cosmology \cite{SFC1,SFC2,SFC3} and group field quantum cosmology \cite{GielenOriti}.
In our case the action will have the same critical points \BD{(that is saddle points)} as the Regge action, which does, in the appropriate regime, reflect the dynamics of the continuum mini-superspace models. Appropriate regimes, where discretization artifacts can be neglected, have been identified in \cite{DGS}. The Regge path integral for these cosmological models has been evaluated in \cite{ADP}, and closely reproduced the continuum mini-superspace results \cite{TurokEtAl}.

A key difference between effective spin foams and Regge path integral is the implementation of a discrete area spectrum into the former class of models. Here we will therefore concentrate on the effects induced by this imposition. Our results indicate that a discrete area spectrum may lead to significant differences for the no-boundary wave functions, affecting the probabilities for the quantum creation of a de Sitter universe.

~\\
This paper is organized as follows:  We will first provide necessary background on effective spin foams (Sec.~\ref{SecEff}), and the  de Sitter mini-superspace model and its discretization using the so-called ball and shell model (Sec.~\ref{SecDisc}). We will  then discuss the issue of configurations with irregular light cone structure (Sec.~\ref{SecIrreg}) and the (cosmological) Regge action (Sec.~\ref{SecRegge} and \ref{SecCrit}).  
Readers familiar with \cite{DGS} or \cite{ADP} can skip Sections  \ref{SecDisc} to  \ref{SecCrit}. 

We will then provide further necessary ingredients for the construction of the cosmological effective spin foam model in Sec.~\ref{SecSpectra} and Sec.~\ref{SecMeas}. Sec.~\ref{SecSum} will summarize the final (ball and shell) models and present the related partition function.

In Sec.~\ref{SecAcc} we will provide some background on non-linear sequence transformations, in particular on the higher order Shanks transform and Wynn's epsilon algorithm. 

Wynn's epsilon algorithm is then applied to compute the partition functions and expectation values for the cosmological effective spin foam models. Sec.~\ref{ResultsShell} will present the results for the cosmological shell model. There we will consider the `Lorentzian regime', that is boundary data which lead to Lorentzian critical points. Sec.~\ref{ResultsBall} will discuss the results for the ball model, where we will consider the `Euclidean regime' — corresponding to boundary data which only has Euclidean (tunneling) saddles. We will for all cases compare the effective spin foam sum with the Regge path integral, and analyze the differences.

We will close with a discussion and outlook on future work in Sec.~\ref{Discussion}.

\section{The cosmological ball and shell model with effective spin foams}

\subsection{Effective spin foam models}\label{SecEff}

Effective spin foam models \cite{EffSF1,EffSF2,EffSF3} are designed to capture key features of the EPRL/FK spin foam models \cite{EPRLFK}, but are much more amenable to numerical calculations. We will leverage this fact here and will provide the first evaluation of  cosmological spin foam amplitudes which involve the summation over a bulk variable, which can be identified as lapse. 

(Effective) spin foams \cite{PerezReview} are defined on triangulations, where all triangles (and possibly all tetrahedra) are decorated with geometric data.  The spin foam partition function sums over these geometric data weighted by a model-specific quantum amplitude. All models have in common that this quantum amplitude leads to the Regge action \cite{Regge}, which provides a very elegant discretization of the Einstein-Hilbert action, in the limit of large quantum numbers.

The two key features of spin foams are the following: Firstly, the spectra for geometric operators, such as area operators are discrete \cite{Smolin,Ashtekar,Conrady}. The spin foam partition function sums over geometries, which are characterized by discrete eigenvalues of the area operators associated to the triangles (and, depending on the model, possibly on additional geometric quantities associated to the tetrahedra).  For spatial triangles the spectral gap is characterized by the dimensionless Barbero-Immirzi parameter, whereas the spectral gap for time-like areas is independent of the Barbero-Immirzi parameter.

These discrete spectra lead to the second feature of spin foams \cite{EffSF1}: an enlargement of the configuration space of geometries \cite{Ryan1}, which can be effectively described as replacing the length metric by the area metric \cite{AreaReggeLimit,JohannaArea,JoseArea}.   Spin foam dynamics imposes shape matching constraints \cite{AreaAngle}, that suppress fluctuations away from length metric configurations \cite{EffSF1,EffSF2,Asante:2022lnp,Han:2023cen}. These constraints are however second class and are therefore only imposed weakly, e.g. via Gaussians peaked on configurations that satisfy the constraints. The Barbero-Immirzi parameter also appears here in characterizing the strength of the suppression,\footnote{The smaller the parameter, the stronger the suppression.} e.g. the as variance of the Gaussians \cite{Ryan3,EffSF1}. 

 As we explain below, due to the symmetry reduction used in the ball and shell models, these shape matching constraints are automatically satisfied, so this second key feature will not play a role here. The main point of the current article is to understand the effects induced by introducing the discrete area spectra into the model. The summation range for the bulk area will mostly involve time-like areas. The choice of Barbero-Immirzi parameter $\gamma$ will therefore have a very limited influence on the results. We will fix it to $\gamma=0.1$, as this value has been previously found to lead to a satisfying implementation of the gravitational equation of motion for triangulations consisting of a few building blocks \cite{EffSF2}.

Whereas the EPRL/FK spin foam amplitudes are defined via a gauge-theoretic reformulation of general relativity \cite{PerezReview}, the effective spin foam models leverage more directly the Regge action. This can be justified through a higher gauge formulation of gravity \cite{HigherGauge}, which leads to a model featuring directly the Regge action \cite{KBF}.  Using directly the Regge action is in particular advantageous if we aim to describe the full Lorentzian spin foam sector.  As mentioned above, the original Lorentzian EPRL/FK models only consider space-like sub-simplices and their extension \cite{Conrady} still faces challenges regarding asymptotics \cite{Simao} and numerical evaluation \cite{NumSFReview}.

This is another reason why effective spin foam models are much easier to handle, and why we are able to implement in this work the first calculation involving a summation over a lapse like variable. 

As we have already mentioned, effective spin foams leverage the Regge action for their construction. This Regge action however use areas as fundamental variables and is therefore known as Area Regge action \cite{AreaRegge,ADH}. This Area Regge action does, in principle, lead to a different dynamics\footnote{Recent work \cite{AreaReggeLimit} has however shown that the continuum limit of the (linearized) Area Regge dynamics leads to (linearized) general relativity, with a correction (second order in the lattice constant) given by the square of the Weyl curvature.} from the original Length Regge action \cite{Regge}.  This difference, again, play no role here due to the symmetry restriction we impose. In fact, we will have a one-to-one transformation between the length and area variables in the model. This allow us to easily impose the discrete spectrum conditions for the areas, see Sections \ref{SecRegge} and \ref{SecSpectra}.

Another simplifying construction step that we will take, is to work with (Lorentzian) flat building blocks and to implement the cosmological constant via a cosmological constant term in the action. This is again straightforward to do in effective spin foam models, but has also been used in simplified versions of the EPRL model \cite{SFC2,BenLambda}.  An alternative procedure is to base the discretization on homogeneously curved building blocks \cite{NewRegge,Lambda}. This reduces discretization artifacts \cite{Improve,Review22}, in particular with regard to the breaking of diffeomorphism symmetry by the discretization \cite{DiffReview1,Broken}. The amplitudes of the resulting spin foam models \cite{Lambda} are however much harder to compute than for the models with flat building blocks, making these models unsuitable for our purposes for now.

\subsection{The discretization}\label{SecDisc}

In this work we aim to construct a numerically accessible model for mini-superspace quantum cosmology, which describes a Lorentzian de Sitter universe. 

We will consider a finite time evolution step in a mini-superspace model given by
\ba\label{minisuper}
{\rm d}s^2\,=\, -N^2(t) {\rm d}t^2+ a^2(t) d\Omega^2 
\ea 
where $d\Omega^2$ is the metric on the unit three-sphere. A given mini-superspace configuration is 
determined by scale factor $a(t)$ and lapse variable $N(t)$ as functions of the time parameter $t$.

Spin foams rely on a discretization of the underlying space time manifold into a piecewise (Minkowski-) flat geometry. We choose a discretization that allows us to describe the discrete time evolution of spatial discretized hypersurfaces, which are as homogeneous as possible, see \cite{DGS,ADP} for more details. 

We will use two models, the so-called ball model and the shell model. Variations of the first model have been used in the context of Euclidean quantum gravity \cite{Hartle1985}, the classical dynamics associated to the shell model has been described in \cite{Collins1973,DGS}.  Recent work \cite{ADP} constructed the Lorentzian Regge path integral describing one time evolution step based on these models. This work \cite{ADP} illustrated that this simple approximation already captured the main key features of the continuum mini-superspace evolution, and in particular modelled the no-boundary wave function.

The so-called ball model approximates the first time evolution step of the quantum de Sitter universe by a subdivided regular 4-polytope. The centre of the 4-polytope corresponds to the beginning of the universe, where the scale factor is vanishing. The outer boundary of the 4-polytope corresponds to a spatial spherical slice of the de Sitter space-time with finite scale factor.   We will consider regular 4-polytopes, whose (three-dimensional) boundary define a triangulation. There are three such 4-polytopes, known as 5-cell, 16-cell and 600-cell. To minimize discretization artifacts \cite{DGS}, we will consider the 600-cell, whose boundary consists of 600 tetrahedra. To implement homogeneity we choose all edge lengths  in this boundary to be the same, and refer to the squared edge length in the boundary as $s_l$.

We subdivide the 600-cell into 600 four-simplices, by placing a vertex in the centre of the 600-cell, and by connecting this central vertex with all the vertices in its boundary. We again set all the lengths of these new edges to be equal. It is however more convenient to use as variable the height square $s_h$ of the four-simplices. For positive $s_l$ and negative $s_h$ the four-simplices satisfy the Lorentzian generalized triangle inequalities, whereas for positive $s_l$ and positive $s_h$ the four-simplices satisfy the Euclidean generalized triangle inequalities. One can understand the height square as a discrete analogue of (minus) the squared lapse parameter.  For the Regge path integral, considered in \cite{ADP}, one integrates over this variable, for the effective spin foam model we will replace this integral by a sum.

The shell model describes a time evolution step between two $600$-cell boundaries, characterized by the squared edge lengths $s_{l_1}$ and $s_{l_1}$, respectively.  Instead of insisting on a triangulation, it is simpler to use as four-dimensional building blocks  frusta\footnote{These frusta can be subdivided into 4-simplices, but this introduces edge additional lengths. In this model we fix these additional edge lengths by demanding that the frusta are piecewise flat building blocks \cite{DGS}. Frusta, which however have a cubic instead of a tetrahedral base, have been also used in \cite{BahrFrusta} in the context of restricted spin foam models.} with tetrahedral base \cite{Collins1973}. We set the length of the edges connecting the two tetrahedra of a given frustum equal, but will use instead of this length the height square $s_h$ of the frusta as dynamical variable. This variable can be again understood as an analogue of (minus) the squared lapse variable. In the Regge path integral we integrate over this variable and for the effective spin foam model we sum over this variable.

In summary, the geometry of the spatial hypersurfaces in the ball and shell model is characterized by $s_l$.  By comparing the three-volume of a 600-cell boundary with the three-volume of a three-sphere in the continuum mini-superspace model, we can deduce the following relation between scale factor $a$ and squared edge length $s_l$ \cite{DGS}, and thus between the scale factor and the area of the triangles in the 600-cell boundary
\ba\label{translation}
s_l \,=\, \nu^2 a^2 \, ,\q\q A\,=\,\frac{\sqrt{3}}{4} s_l\,=\, \frac{\sqrt{3}}{4} \nu^2  a^2 \, ,
\ea
where $\nu \approx 0.654$. 

The lapse parameter in (\ref{minisuper})  is replaced by the squared height $s_h$. This squared height can be also translated into the area square $\mathbb{A}^{\rm Ball}_{\rm blk} $ of the bulk triangles or area square $\mathbb{A}^{\rm Shell}_{\rm blk}$ of the bulk trapeziums in the ball and shell model, respectively
\ba
  \mathbb{A}^{\rm Ball}_{\rm blk} \,=\,\frac{\sqrt{s_l}}{4\sqrt{2}}\,\,(s_l+8 s_h) \, ,\q\q
 \mathbb{A}^{\rm Shell}_{\rm blk} \,=\,\frac{\sqrt{s_{l_1}}+ \sqrt{s_{l_2}  }}{4\sqrt{2}}\, \left((\sqrt{s_{l_2}}-\sqrt{s_{l_1}})^2+8 s_h\right) \q .
\ea
Note that these area squares can be negative, null, or positive.

\subsection{Regular and irregular light cone structure}\label{SecIrreg}

Even though the models we consider are quite simple, they feature configurations with an irregular light cone structure. 

In a smooth Lorentzian manifold a point comes with a regular light cone structure, if in a small ball around this point there are exactly two light cones attached to this point. For piecewise Minkowski-flat geometries, which we consider here, one can formulate a set of conditions,\footnote{In \cite{LollLC} these are dubbed local causality condition. However one can argue that even to points violating these conditions, one can assign a well defined future and past. We thus prefer the notion of irregular light cone structure.} which ensure a regular light cone structure \cite{LollLC,ADP}. 

Of particular importance is one such condition, which we will refer to as hinge regularity. Configurations which violate this condition feature imaginary terms in the Regge action, which comes with a sign ambiguity. This sign ambiguity can be understood as a branch cut for the complexified Regge action \cite{ADP}. 

Hinge regularity is of particular importance as it is a condition which involves the deficit angles, which are a key ingredient for the Regge action.  The deficit angles are attached to the building blocks of co-dimension two, known as hinges. To define the deficit angle one projects out from all building blocks attached at the hinge the hinge itself. This leads to a two-dimensional piecewise flat manifold given by two-dimensional building blocks $b$, which all share the vertex $v$ resulting from the hinge. The deficit angle is defined as the difference between the flat angle and the sum of the two-dimensional angles $\theta^b_v$ at $v$ in the building blocks $b$. 

If the hinge is time-like the two-dimensional building blocks $b$ resulting from the projection will all be space-like and thus the angles $\theta^b_v$ will be Euclidean and the flat angle is given by $2\pi$. 

If the hinge is space-like the two-dimensional building blocks $b$ will be time-like and we have to deal with Lorentzian angles $\theta^b_v$.  These Lorentzian angles, as defined in e.g \cite{Sorkin1974,Sorkin2019, Ding2021,ADP} will in general include a real part and an imaginary part $\pm \imath k \pi/2$, which comes with a sign ambiguity. Here $k$ denotes the number of light rays included in the wedge described by $\theta^b_v$. Depending on the sign ambiguity, the Lorentzian flat angle is given by $\pm 2\pi \imath$.  

Hinge regularity is violated if the union of the wedges described by the $\theta^b_v$ does not include exactly four light rays, or two light cones. In case of such a violation the deficit angle includes an imaginary part with a sign ambiguity. 

Note that for hinge regular configurations the choice of sign does not affect the deficit angles.

The Regge action  \cite{Regge,Sorkin1974} is given by the sum of the deficit angles weighted by the volume (here area) of the hinges.\footnote{In case the triangulation has a boundary one has a (Gibbons-Hawking-York) boundary term \cite{HartleSorkin}. This boundary term has the same structure as the bulk term, the only difference is that the deficit angles are replaced by extrinsic curvature angles. The latter are computed in the same way as the deficit angles, one only needs to replace the flat angle with the flat half-angle, or some other choice if one considers corners.}   Hinge regularity violations lead thus to imaginary parts in the Regge action. Depending on the sign of these imaginary parts, these can lead either to a suppression or enhancement of hinge regularity violating configurations in the path integral. 

Both choices of sign for the Lorentzian action and the Euclidean action arise from a notion of complex Regge action \cite{ADP}. The sign ambiguity can then be associated to a branch cut, which goes along hinge regularity violating configurations. For the Lorentzian path integral we then have to decide whether to include such hinge irregular configurations. If we do include such configurations, we have to choose a side for each branch cut. 

With the methods presented here we can consider each of the options. \cite{ADP} has however found that including irregular configurations for the ball model does lead to a real amplitude, as one also finds in the continuum. We will therefore include irregular configurations and choose the suppressing side, as has also be done in \cite{ADP}.  For some cases we will provide a comparison to the choice where one does not include the irregular configurations.

\BD{As discussed above the ball model allows us to obtain the no-boundary wave function from the path integral.  \cite{ADP} showed that choosing the suppressing side of the branch cut will lead to an approximation of the Vilenkin version of the no-boundary wave function.   With this choice we obtain an exponential suppression of the wave function scaling with $\exp(-\frac c{\hbar\Lambda})$ (with $c>0$ a constant factor), as expected from quantum mechanical tunnelling.  This Vilenkin version of the no-boundary wave function has also been obtained from a Lorentzian mini-superspace model in \cite{TurokEtAl} and a Regge path integral in \cite{ADP}.  The no-boundary wave function from the Regge path integral does approximate the  one obtained  from the continuum mini-superspace path integral surprisingly well. A key question we will consider here, is how the discrete area spectra do the results obtained from the Regge path integral.

Note that one can also choose the enhancing side of the branch cut. This has been done in the recent work \cite{Dittrich:2024awu}, in order to compute the entropy for de Sitter space via a simplicial Lorentzian path integral. To obtain the de Sitter entropy one indeed needs an exponentially enhanced result scaling with $\exp(+\frac c{\hbar\Lambda})$.  The corresponding version of the no-boundary wave function is the Hartle-Hawking one. 

Although the light cone irregularities and therefore the choice of branch cut side appear to be a discretization artifact, there is a similar ambiguity of integration contour in the continuum, as illustrated by the fact that \cite{HartleEtAl} obtained the Hartle-Hawking version of the no-boundary wave function from a Lorentzian mini-superspace model. The difference between \cite{TurokEtAl} and \cite{HartleEtAl} lies in the choice of how the integration contour navigates an essential singularity in the continuum mini-superspace action at vanishing lapse. We refer the reader to \cite{Dittrich:2024awu} for a more in-depth discussion of this issue.
}

\subsection{The  Regge action}\label{SecRegge}

The effective spin foam models rely on the Regge action \cite{Regge}, which provides a discretization of the Einstein-Hilbert action.  This discretization is based on triangulations, and in (length) Regge calculus one uses lengths attached to the edges of this triangulation as basic variables. In the case of effective spin foams, one uses instead areas as basic variables, and thus the area Regge action \cite{AreaRegge, ADH}, which is accompanied by gluing or shape matching constraints \cite{AreaAngle, EffSF1,EffSF3}. But with the symmetry reduction employed here,  we will have the same number of length and area variables, and length and area Regge action will lead to the same equations of motion. That is, as mentioned above, the shape matching constraints are automatically satisfied. 

We can thus start with the length Regge action, which has been already computed for the cosmological model we will consider here \cite{DGS,ADP}.  We can then implement a variable transformation to areas, which will allow us to impose that these have a discrete spectrum, as in effective spin foam models. 

Using the notation of \cite{ADP} the Regge action for the ball and shell model have the following form 
\ba\label{Action1}
\pm\imath \,(8\pi G) S^{\pm}_{\rm Ball}&=& n_e \sqrt{\!\!{}_{{}_\pm} \mathbb{A}_{\rm blk}} \,\delta^{\pm}_{\rm blk} + n_t \sqrt{\!\!{}_{{}_\pm}\, \mathbb{A}_{\rm bdry}} \,\delta^{\pm}_{\rm bdry} -\Lambda n_\tau \sqrt{\!\!{}_{{}_\pm} \mathbb{V}_\sigma} \, \,\,,\nn\\
\pm\imath \,(8\pi G) S^\pm_{\rm Shell}&=& n_e \sqrt{ \mathbb{A}_{\rm blk}} \,\delta^\pm_{\rm blk} + n_t  \left(\sqrt{\!\!{}_{{}_\pm} \mathbb{A}_{\rm bdry_1}} \, \,\delta^\pm_{\rm bdry_1}  + \sqrt{\!\!{}_{{}_\pm} \mathbb{A}_{\rm bdry_2}} \, \,\delta^\pm_{\rm bdry_2}  \right)   -\Lambda n_\tau \sqrt{\!\!{}_{{}_\pm}\mathbb{V}_{\text{4-frust}}}  \,\,\, .
\ea
Here $n_e=720$, $n_t=1200$ and $n_\tau=600$ denote the number of edges, triangles and tetrahedra in the boundary of the 600-cell. (The action does generalize to the 5-cell and 16-cell if one replaces these numbers appropriately.)
Note that the number of edges in the boundary of the 600-cell is equal the number of bulk triangles.  We collected the expressions for the complex deficit angles $\delta_{\rm blk}$, the boundary angles $\delta_{\rm bdry}$, the signed area squares $\mathbb{A}_{\rm blk}$ and  $\mathbb{A}_{\rm bdry}$ and the signed four-volume squares $\mathbb{V}_\sigma$ in Appendix \ref{AppA}.

In (\ref{Action1}) we introduced two different ways to express the complex Regge action \cite{ADP}. These differ in the evaluation of the square root and the logarithm (which appears in the deficit angles $\delta^{\pm}$, see Appendix \ref{AppA}) along the principal branch cut, that is the negative real axis, for these functions. The branch cut values are defined by  
 $\sqrt{\!\!{}_{{}_+}-1}=+\imath,\,\,\sqrt{\!\!{}_{{}_-}-1}=-\imath$ and $\log_+(-1)=+\imath\pi,\,\, \log_-(-1)=-\imath\pi$.
 
 The action $S^{+}$ and $S^{-}$ evaluate to the same function for Lorentzian data which satisfy the hinge  regularity condition. This is the case if $s_h<-\tfrac{1}{8}s_l$ for the ball model and if $s_h<-\tfrac{1}{8}(\sqrt{s_{l_2}}-\sqrt{s_{l_1}})$ for the shell model.  Starting from this regular Lorentzian region we can analytically continue this action\footnote{See \cite{ADP} for an extensive discussion on the analytical continuation of this action and the resulting Riemann surface.} to negative values of $s_h$, which violate these inequalities and describe hinge irregular configurations. As there is a branching point at $s_h=-\tfrac{1}{8}s_l$ (or $s_h=-\tfrac{1}{8}(\sqrt{s_{l_2}}-\sqrt{s_{l_1}})$  for the shell model), we find different actions when we analytically continue along $s_h=x+\imath \varepsilon$ or $s_h=x-\imath \varepsilon$ with $x$ negative.  Going along $s_h=x+\imath \varepsilon$ we find $S^{+}$, which for the hinge violating configurations has a negative imaginary term, and going along $s_h=x-\imath \varepsilon$ we find $S^{-}$, which for the hinge violating configurations has a positive imaginary term. As in \cite{ADP}, we will work with the choice which suppresses hinge violating configurations, that is $S^-$.

In a Lorentzian (non-degenerate) triangulation we have $s_h<0$. This translates into negative or positive bulk area squares. Negative bulk area squares $\mathbb{A}_{\rm blk}<0$ describe hinge regular configuration, the null case $\mathbb{A}_{\rm blk}=0$ leads to a branching point for the Regge action. 

For the ball model, the hinge irregular Lorentzian regime includes bulk areas with $0<\mathbb{A}_{\rm blk}<\tfrac{1}{6} \mathbb{A}_{\rm bdry}$, for the shell model the Lorentzian regime extends to $\mathbb{A}_{\rm blk}<\tfrac{1}{6} (\sqrt{\mathbb{A}_{{\rm bdry}_2}}-\sqrt{\mathbb{A}_{{\rm bdry}_1}})^2$.    These configurations include a further branching point $\mathbb{A}_{\rm blk}=\tfrac{1}{9} \mathbb{A}_{\rm bdry}$, respectively $\mathbb{A}_{\rm blk}=\tfrac{1}{6} (\sqrt{\mathbb{A}_{{\rm bdry}_2}}-\sqrt{\mathbb{A}_{{\rm bdry}_1}})^2$, where the three-dimensional bulk building blocks are null. 

The imaginary parts of $S^-$ are given by
\ba
8\pi G\,\, {\rm Im}(S^-_{\rm Ball})&=& \begin{cases}
   n_e 2\pi \sqrt{ \mathbb{A}_{\rm blk}} \q \q\q\q\q\q\q\q\;
 \text{for} \,\, 0<\mathbb{A}_{\rm blk}<\tfrac{1}{9}\mathbb{A}_{\rm bdry} \\
  -n_e \pi \sqrt{ \mathbb{A}_{\rm blk}}+ n_t \pi  \sqrt{ \mathbb{A}_{\rm bdry}}
 \q \text{for} \,\, \tfrac{1}{9}\mathbb{A}_{\rm bdry}<\mathbb{A}_{\rm blk}<\tfrac{1}{6}\mathbb{A}_{\rm bdry}
\end{cases} \, ,\nn\\
8\pi G\,\, {\rm Im}( S^-_{\rm Shell})&=& \begin{cases}
   n_e 2\pi \sqrt{ \mathbb{A}_{\rm blk}} \q \q\q\q\q\q\q\q\; 
 \text{for} \,\, 0<\mathbb{A}_{\rm blk}<\tfrac{1}{9}(\sqrt{\mathbb{A}_{{\rm bdry}_2}}-\sqrt{\mathbb{A}_{{\rm bdry}_1}})^2\\
  -3n_e \pi \sqrt{ \mathbb{A}_{\rm blk}}+ n_t \pi  (\sqrt{ \mathbb{A}_{{\rm bdry }_2}}  - \sqrt{ \mathbb{A}_{{\rm bdry}_1}} )
 \\ \q\q\q\q\q\q\q \text{for} \,\, \tfrac{1}{9}(\sqrt{\mathbb{A}_{{\rm bdry}_2}}-\sqrt{\mathbb{A}_{{\rm bdry}_1}})^2<\mathbb{A}_{\rm blk}<\tfrac{1}{6}(\sqrt{\mathbb{A}_{{\rm bdry}_2}}-\sqrt{\mathbb{A}_{{\rm bdry}_1}})^2
\end{cases} \, ,
\ea
and are positive for the entire hinge irregular regime.
 
The suppressing effect by the imaginary part of the action can be used to avoid summation over the range of values where the imaginary part of the action is very large.

\subsection{Critical points of the action}\label{SecCrit}

The existence of critical points for the action in the ball and shell model has been analyzed in \cite{DGS}. We will summarize the main points here and will start with the ball model.  Here one does not find a critical point for negative, that is Lorentzian, values of $s_h$ (or the bulk area) as long as the boundary length square $s_l$ (or $A=\sqrt{\mathbb{A}_{\rm bdry}}$) is below a certain threshold value $s_{\rm thresh1}$. Instead, one finds a critical point for positive, that is Euclidean values of $s_h$. This feature mimics the behaviour of the mini-superspace model, where one finds Euclidean critical values,\footnote{In the continuum model one finds two Euclidean critical values for positive imaginary lapse. These two values correspond to an evolution which remains in one hemi-sphere of the de Sitter sphere and an evolution which crosses the equator and therefore involves both hemi-spheres. Using only one time step in the discrete model we only see the first case, to see the second solution one needs at least two time steps.} if the boundary values for the scale factor are both smaller than $a_\Lambda=\sqrt{3/\Lambda}$.  In the continuum, the Hamilton-Jacobi function, that is the action evaluated on the critical points, shows a monotonous behaviour. In the discrete, this is only the case for $s_l<s_{\rm thresh2}\approx 1.75/\Lambda$ (for the 600-cell), which translates to a scale factor $a_{\rm thresh2}\approx \sqrt{4.08/\Lambda}$.  We consider the behaviour in the regime above this threshold as discretization artifact, and will therefore only consider boundary values below this threshold. 

The shell model allows to model the behaviour of the continuum model, when both boundary scale factors are above the value $a_\Lambda=\sqrt{3/\Lambda}$. One then finds critical points in the Lorentzian regime.\footnote{In the continuum one again finds two solutions corresponding to the solution not crossing or crossing the equator of the Lorentzian de Sitter space. In the discrete (with only one time step) we only find the solution which does not cross the equator.} For the shell model we will therefore consider boundary values which lead to Lorentzian critical values.

\subsection{Discrete area spectra}\label{SecSpectra}

In (Length) Quantum Regge calculus the path integral is defined as a continuous integral over the length of the bulk edges of the triangulation. In spin foams one instead sums over discrete values assigned to the areas of the bulk triangles.   These discrete values are determined from the discrete spectra for the loop quantum gravity area operator.  For space-like areas (in a space-like tetrahedron) the spectrum is given by \cite{Smolin,Ashtekar}
\ba\label{SpecS}
\mathbb{A}_{\rm space-like} = \ell_{\rm P}^4 \gamma^2 j(j+1)  \sim   \ell_{\rm Planck}^2 \gamma^2 j^2 \, , \q j=\tfrac{1}{2},1,\tfrac{3}{2},2,\tfrac{5}{2}.\ldots \q .
\ea
Here $\ell_{\rm P}^2=8\pi G\hbar$ is the Planck length squared  and $\gamma$ is the dimensionless Barbero-Immirzi parameter. We choose this parameter as $\gamma=0.1$, which is within the range where one can expect the recovery of the Regge equations of motion from spin foams \cite{EffSF2}. The spectrum for the area becomes equidistant for large $j$. For simplicity,\footnote{One actually has  $\mathbb{A}_{\rm space-like}=\ell_{\rm Planck}^4 \gamma^2 j(j+1)$ only for space-like areas in a space-like tetrahedron. For space-like areas in a time-like tetrahedron one rather has $\mathbb{A}_{\rm space-like}=\ell_{\rm Planck}^4 \gamma^2 j(j-1)$ and the condition $j\geq 1$ \cite{Conrady}. The equidistant spectrum is a compromise, where we do not need to distinguish between these two types of space-like areas.} we will adopt this equidistant spectrum and include only integer $j$'s. (The latter corresponds to a definition of loop quantum gravity as $\text{SO}(3)$ gauge theory instead of a $\text{SU}(2)$ gauge theory.)

The spectrum for time-like areas is given by \cite{Conrady}
\ba\label{SpecT}
\mathbb{A}_{\rm time-like} = \ell_{\rm Planck}^4 (n/2)^2  \,, \q n=0,1,2,3,\ldots  \q .
\ea
Thus the time-like spectrum does {\it not} depend on the Barbero-Immirzi parameter $\gamma$. Similarly to the space-like case, we will restrict to $n\in 2\mathbb{N}$.

Adopting these spectra for the area squares of a single triangle (or trapezium) in our symmetry reduced model will however not lead to sensible results. The reason for this is that the change from one spectral value to the next spectral value induces a very large change in the action. That is the number of spectral values over a few oscillations of the phase of the amplitude $\exp(\imath S)$ is very low, even around the saddle point,  and we can therefore not expect a semi-classical regime. See Fig.~\ref{FigShell1} for an example.

This issue is caused by the symmetry reduction. Indeed, assume we change the area of a bulk triangle by the spectral gap value  $(\gamma) \ell_{\rm P}^2$. Due to our symmetry requirements,  the total area will change by $n_e (\gamma) \ell_{\rm P}^2$, with $n_e=720$, leading to a rather large change in the action. 

Thus we adjust the spectrum to the symmetry reduction: we divide the spectral gaps for the bulk triangles by the number of bulk triangles (or bulk trapeziums) $n_e$ and the spectral gap for the boundary triangles by the number of boundary triangles $n_t=1200$.  Thus we have for the bulk triangles
\ba\label{BulkSp}
\mathbb{A}_{\rm space-like,bulk} =\frac{  \ell_{\rm Planck}^2}{n_e^2} \gamma^2 j^2 \, , \q j=1,,2,\ldots \q,\q\q
\mathbb{A}_{\rm time-like,bulk} =\frac{  \ell_{\rm Planck}^2}{n_e^2}  n^2 \, , \q n=1,2,\ldots \q . \q\q
\ea
 This choice can be justified by the following consideration: Consider a homogeneous configurations with sufficiently large areas. Changing the area of $X$ of the $n_e$ bulk triangles by the spectral gap value  will still give a configuration which is almost homogeneous and whose action can be approximated by assuming a homogeneous configuration with a suitable adjusted area value. With $X$ running from $1$ to $n_e$ we obtain that the total area, that is the sum of all bulk areas, has changed by $n_e$ times the minimal allowed spectral value. 
To implement that the total area can change by the spectral gap value, we therefore adjust the spectrum of a single bulk triangle (trapezium) by dividing the area spectral gaps by the number of bulk triangles $n_e$. Similarly, we adjust the spectrum for a single boundary triangle by dividing the area spectral gap by the number of boundary triangles $n_t$.

Of course, we have made here many simplifying assumptions. These will have to be confirmed or corrected  by future work which will investigate constructions with less symmetry reductions. These new insights might in particular affect the spectrum for the regime where the total area is relatively small.

A similar need for the adjustment of spectral gaps appears  also in Loop Quantum Cosmology \cite{BojowaldLivRev, Mubar} and the task to fully justify these choices from the full theory is still open \cite{LQCfromFull}.

To be able to trace back how the results change if we replace the continuous Regge path integral by the effective spin foam sum we will also consider a further refinement of the area spectra. That is, we divide the spectral gap by a refinement parameter $R_{\rm ref}$, and compare the spin foam sum using the spectrum (\ref{BulkSp}), with the spin foam sum using a refined spectrum and with the Regge path integral, or in other words the sum with an infinitely refined spectrum.

\subsection{The path integral measure}\label{SecMeas}

The second ingredient we have to fix is the measure term for the spin foam summation. As we wish to primarily study the difference between doing a (spin foam) summation and doing a (Regge) path integral, we will adopt the measure from the (Regge) path integral \cite{ADP}. This Regge measure was derived from the continuum path integral constructed in \cite{TurokEtAl} and led to a very good agreement between the Regge path integral and continuum path integral.


The (dimensionless) measure for the Regge path integral in the ball model is given by \cite{ADP}   
\ba
\mu_{\rm Ball}(s_h)\bd s_h&=&\frac{(-1)^{\frac{3}{4}}}{\ell_P}\frac{s_l^{\frac{1}{4}}}{(-s_h)^{\frac{3}{4}}}\bd s_h
\ea
where the integral is over negative $s_h$ (with the integral oriented to go from $s_h=0$ to $s_h\rightarrow -\infty$ ).  Transforming this measure to area variables gives for the ball model
\ba\label{MeasB}
\mu_{\rm Ball}^t( B_{t})\bd B_t \,=\,\frac{(-1)^{-\frac{1}{4}}}{\ell_P} \frac{4 \times 2^{\frac{1}{4}} B_t}{\left(\tfrac{\mathbb{A}_{\rm bdry}}{3}+2 B_t^2\right)^{\tfrac{3}{4}}} \bd B_t \, ,\q\q
\mu^s_{\rm Ball}( B_{s})\bd B_s\,=\, \frac{(-1)^{-\frac{1}{4}}}{ \ell_P}  \frac{4 \times 2^{\frac{1}{4}} B_s}{\left(\tfrac{\mathbb{A}_{\rm bdry}}{3}-2 B_s^2\right)^{\tfrac{3}{4}}} \bd B_s\q\q .
\ea
Here we parametrize the time-like range of the bulk area by $-B_t^2=\mathbb{A}_{\rm blk}$ and the space-like range of the bulk area by $B_s^2=\mathbb{A}_{\rm blk}$ with $B_t,B_s >0$. 
For the time-like range we integrate from $B_t=0$ to $B_t\rightarrow \infty$, and for the space-like range from $B_s=0$ to 
$B_s= \sqrt{\mathbb{A}_{\rm bdry}}/\sqrt{6}$. 

We define the measure for  the shell model by replacing in $\mu_{\rm Ball}$ the boundary variable $\mathbb{A}_{\rm bdry}$ by the average $\mathbb{A}_{\rm bdry}\rightarrow \tfrac{1}{2}(\mathbb{A}_{{\rm bdry }_1}+\mathbb{A}_{{\rm bdry 2}_2})$.

\subsection{The spin foam sum}\label{SecSum}
The final definition of our cosmological spin foam sum for the ball model is as follows:
\ba\label{ESFSum}
Z_{\rm Ball}(\mathbb{A}_{\rm bdry} )&=&  \frac{  \gamma  \ell_{\rm P}^2}{R_{\rm ref} n_e}    \sum_{j_B=1}^{j_B^{\rm max}}  \mu^s_{\rm Ball}\left(  \frac{  \gamma  \ell_{\rm P}^2  j_B}{R_{\rm ref} n_e}  \right)  \exp\left( \frac{\imath}{\hbar} S_{\rm Ball}\left(   \mathbb{A}_{\rm bdry}  \,,\, \left(\frac{ \gamma  \ell_{\rm P}^2  j_B}{R_{\rm ref} n_e} \right)^2\right) \right)\,\,+ \nn\\
&&  \frac{   \ell_{\rm P}^2}{R_{\rm ref} n_e}    \sum_{n_B=1}^{\infty}  \mu^t_{\rm Ball}\left(    \frac{ \ell_{\rm P}^2 n_B}{R_{\rm ref} n_e}  \right)  \exp\left( \frac{\imath}{\hbar} S_{\rm Ball}\left(   \mathbb{A}_{\rm bdry}  \,,\, \left(   \frac{\ell_{\rm P}^2 n_B}{R_{\rm ref} n_e} \right)^2\right) \right) \q 
\ea
where $j_B^{\rm max}=\left[\left[\frac{ R_{\rm ref} \sqrt{\mathbb{A}_{\rm bdry} }}{\sqrt{6}}\frac{n_e}{\gamma}  \right]\right]$, and $\left[\left[x\right]\right]$ defines the integer part of $x$. We will later use $R_{\rm ref}=1$, that is the spectrum (\ref{BulkSp}), as well as a refinement factor of $R_{\rm ref}=10$.

The partition function for the shell is of the same form as (\ref{ESFSum}). But it now depends on two boundary values $\mathbb{A}_{{\rm bdry}_1}$ and $\mathbb{A}_{{\rm bdry}_2}$ and one has  $j_B^{\rm max}=\left[\left[\frac{ R_{\rm ref} (\sqrt{\mathbb{A}_{{\rm bdry}_2} }- \sqrt{\mathbb{A}_{{\rm bdry}_1} }  )    }{\sqrt{6}}\frac{n_e}{\gamma}  \right]\right]$.  We will see that the differences between spin foam sum and Regge path integrals will be minimal, hence we will only consider $R_{\rm ref}=1$.

Apart from the partition function, we will compute the expectation value of  $\mathbb{A}_{\rm blk}$. To this end we insert $B_s^2$ and $-B_t^2$ into the sum over $j_B$ and $n_B$, respectively.

~\\
The second sum in (\ref{ESFSum}) includes infinitely many terms and is converging only very slowly (see Fig.~\ref{FigAcc1}). It even diverges if we insert $-B_t^2$ for the computation of expectation values (see Fig.~\ref{FigAcc2}). But it turns out that so-called acceleration techniques for series convergence work very well in this case, and allow also the evaluation of (absolutely) diverging sums arising for the computation of expectation values.

\section{Acceleration operators for series convergence}\label{SecAcc}

\begin{figure}[ht!]
\begin{picture}(500,150)
\put(0,2){ \includegraphics[scale=0.6]{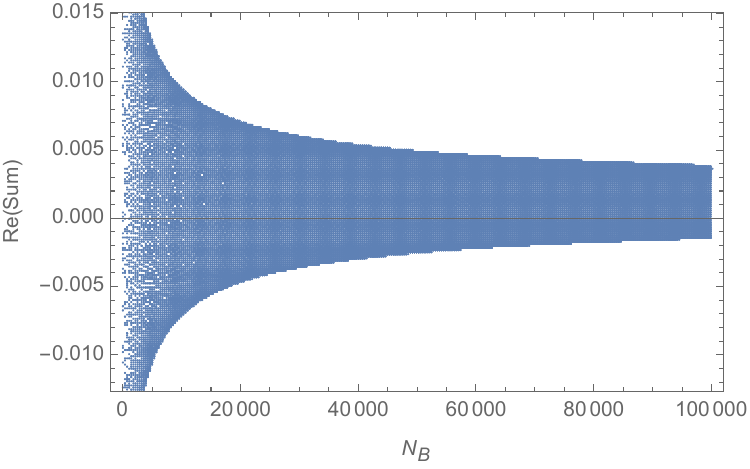} }
\put(260,2){ \includegraphics[scale=0.6]{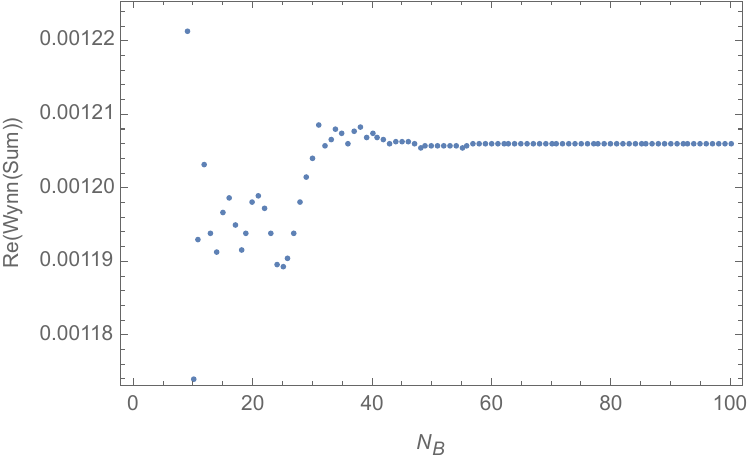} }
%
\end{picture}
\caption{ \footnotesize 
The plot on the left panel shows the partial sums over time-like areas with cut-off $N_B$ for the ball model. We sum all $n_B$ with $n_B\leq N_B$.  (The parameters for the spin foam sum area $\Lambda=0.2\ell_P^{-2}$ and $\sqrt{\mathbb{A}_{\rm bdry} }\approx 0.033\ell_P^2$.)
 On the right panel we show the series resulting from applying Wynn's epsilon algorithm (which will be explained in the main text) to the series defined by the first 100 partial sums shown on the left. The resulting series has a highly accelerated convergence. The maximal relative error (defined in (\ref{error})) is of the order of $10^{-11}$.
 \label{FigAcc1}}
\end{figure}

\begin{figure}[ht!]
\begin{picture}(500,150)
\put(0,2){ \includegraphics[scale=0.6]{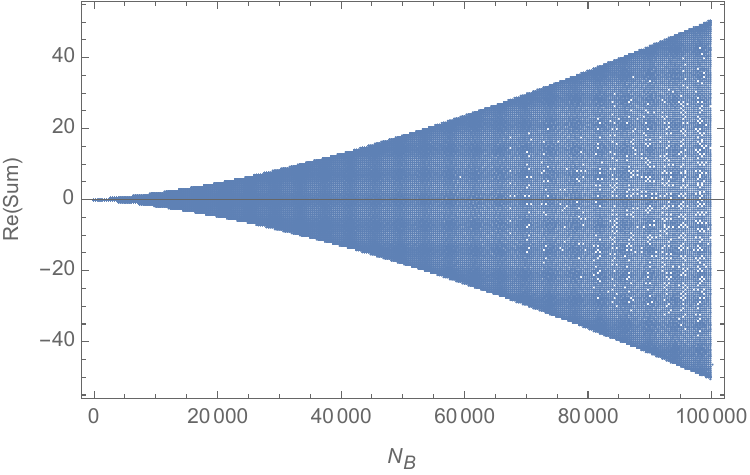} }
\put(260,2){ \includegraphics[scale=0.6]{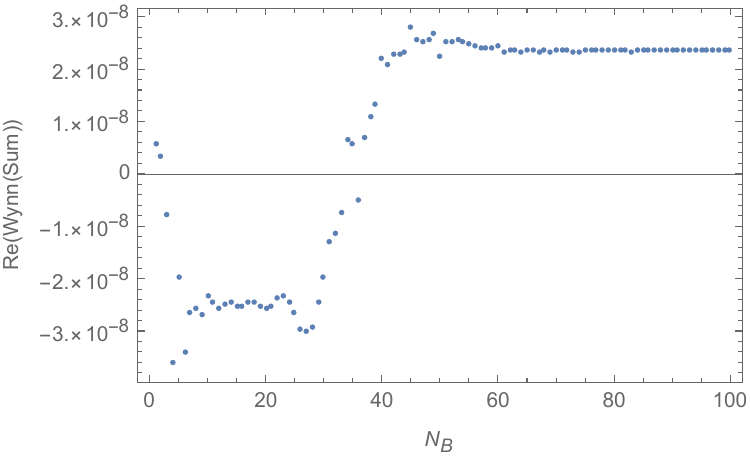} }
%
\end{picture}
\caption{ \footnotesize 
The plot on the left panel shows the partial sums over time-like areas for the computation of the expectation values. That is compared to the sums shown in Fig.~\ref{FigAcc2} we insert a term proportional to $n_B^2$. As before we sum over all positive values $n_B\leq N_B$.  The right panel shows Wynn's epsilon algorithm applied to the series defined by the first 100 partial sums.  This series shows quite a fast convergence. Note that the (anti-) limit is a very small number, which we found is typical for the computation of the expectation values in the ball model. The maximal relative error (defined in (\ref{error})) is of the order of $10^{-8}$. 
\label{FigAcc2}}
\end{figure}

As we explained in the previous sections we have to evaluate oscillating and slowly converging, or even diverging, sums.  Direct summation would be very cumbersome, e.g. Fig.~\ref{FigAcc1} shows an example for an effective spin foam sum for the ball model. We show the partial sums with up to $10^5$ terms, but the result still oscillates with an amplitude which is larger than the absolute value of the limit value.  For the evaluation of expectation values we even encounter diverging sums, see Fig.~\ref{FigAcc2}. Given that we aim to evaluate a large number of such sums, this process of direct summation is  not practical.

Fortunately, there exist a well-developed theory of acceleration operators, also known as non-linear sequence transformations, see e.g. \cite{WenigerReview}. Such transformations and algorithms, are constructed with the aim to accelerate the convergence of slowly converging series, and can also be used to define limit values to divergent series.  We will see that these limit values lead to physically reasonable results: e.g. for the computation of expectation values we rely on such limit values, and in most cases  the expectation values we compute will approximate well the classical solutions.

The non-linear sequence transformations can be applied to compute the limit of sums or integrals with infinite summation or integration range, respectively. To treat sums we form a series from the partial sums $S_k,k=0,1,\ldots$
\ba\label{Accsum}
S_k= \sum_{n=1}^{C_{\rm min }+k C_{\rm step}} f(n) \q .
\ea
Here one can choose an arbitrary minimal cut-off  $C_{\rm min}$ for the sum, so that  $S_0$ represent the sum of $C_{\rm min}$ terms. $C_{\rm step}$ is the step size for probing the partial sums. We found that choosing $C_{\rm step}=1$ leads often to the best results. 

For the application to integrals we define 
\ba\label{Accint}
S_k= \int_{x_0}^{C_{\rm min}+ k\times C_{\rm step}} f(x) \,dx \q .
\ea
$C_{\rm step}$ should be chosen such that the $S_k$ probes the (largest frequency) oscillations of the integral $S(y)=\int_{x_0}^y f(x) dx$, i.e. there should be several $S_k$ for each period.

In the following we will explain how non-linear sequence transformations can accelerate sequence convergence and briefly sketch the construction of Wynn's epsilon algorithm, which we will use to compute the cosmological Regge path integrals and effective spin foam sums.

\subsection{Shanks transform and Wynn's epsilon algorithm}

To begin with we will discuss the Shanks transform. The Shanks transform was derived by Schmidt \cite{Schmidt} in 1941, and rediscovered and popularized by Shanks \cite{Shanks} in 1955.  To motivate this transform, we consider a model series $\{S_k\}_k$ which we assume to be of the form 
\ba\label{S3}
S_k= S+ \alpha t^k \q .
\ea
The series converges for $|t|<1$ to its limit value $S$.  The term $\alpha t^k$ is called a transient.  In case that $|t|>1$ we have a divergent series. One can nevertheless define $S$ as limit value (also known as anti-limit) of this series.

Now if our series $\{S_k\}^M_{k=1}$ is (approximately) of the form (\ref{S3}), we can (approximately) determine $S$ from three consecutive values $S_{k},S_{k+1}$ and $S_{k+2}$. Applying (\ref{S3}) to these three values leads to three equations, which can be solved for the three unknowns $S,\alpha$ and $t$. The solution for $S$ defines the first order Shanks transform $\{T^1_k\}^{M-2}_{k=1}$, given by
\ba\label{Aitken}
T^1_k \equiv  T^1(S_k,S_{k+1},S_{k+2})= S_{k} - \frac{ (S_{k+1}-S_{k})^2}{(S_{k+2}-S_{k+1})-(S_{k+1}-S_{k})} \q .
\ea

 If the original sequence converges, the value of $T^1_k$ approximates $S$ increasingly better as $k$ is increased (for high enough values) and does so faster than $S_k$. It is in this sense that convergence has been ``accelerated''.

In the case that the series $\{S_k\}_k$ has $M$ distinct transients, i.e. is of the form
\ba\label{S5}
S_k=S+\sum_{m=1}^M \alpha_m t_m^k
\ea
with $1>|t_1|>|t_2|>\cdots |t_M|$, the first order Shanks transform eliminates the dominant transient term $\alpha_1 t_1^k$ \cite{WenigerReview}.   This suggest to iterate the Shanks transform in order to eliminate all the transients.   


An alternative procedure to eliminate a larger number of transients from a series, which is (approximately) of the form (\ref{S5}) is to apply an $M$-th order Shanks transform.  Here one applies equation (\ref{S5}) to the $(2M+1)$ consecutive values $S_k,S_{k+1}, \ldots, S_{k+2M}$ to determine the $(2M+1)$ unknowns $\{\alpha_m,t_m\}_{m=1}^M$ and $S$.  The solution for $S$ can be, via Cramer's rule, expressed via a ratio of determinants, see e.g. \cite{WenigerReview}. Thus the $M$-th order Shanks transform
\ba
T^M_k\equiv T^M(S_k, \cdots, S_{k+2M})
\ea
returns a series of length $N-2M$, if the initial series is of length $N$. Note that the same holds for the $M$-th iteration of the first order Shanks transform.

Computing determinants is numerically expensive. Fortunately, Wynn's epsilon algorithm allows the computation of the higher order Shanks transform via a simple recursive scheme. To this end one defines
\ba\label{Wynn1}
\epsilon_k^{(-1)}&=&0 \,\, ,\q\q \epsilon_k^{(0)}=S_k \nn\\
\epsilon_k^{(m+1)}&=& \epsilon^{(m-1)}_{k+1} +\frac{1}{ \epsilon^{(m)}_{k+1}-\epsilon^{(m)}_{k} }\,\,, \q\q k,m \leq 1        \q .
\ea

Wynn \cite{Wynn} could show that the epsilon values with even superscripts $\epsilon_k^{2m}$ reproduce the $m$-th order Shank transformations
\ba\label{Wynn2}
\epsilon^{(2m)}_{k}&=& T^m_k     \q ,
\ea
whereas the epsilon values with odd superscripts are auxiliary quantities.

Thus, in order to compute a high order Shanks transform one can proceed by repeatedly using the second equation in (\ref{Wynn1}), which allows to determine the values $\{\epsilon^{(m+1)}_{k}\}_{k=1}^{N-2}$ as a function of $\{\epsilon^{(m)}_{k}\}_{k=1}^{N-1}$ and $\{\epsilon^{(m-1)}_{k}\}_{k=1}^{N}$.  But in fact, it is not really necessary to proceed by computing  the values $\epsilon^{(m)}_{k}$ for all $k$ for a fixed $m$.  

It is more efficient \cite{Wynn2}, to increase $k$ in each iteration step by one, and to compute for each fixed $k$, the values $\epsilon_{k-m}^{(m)}$.  Wynn's moving lozenge technique implements \cite{Wynn2} such an iterative scheme, see  \cite{WenigerReview} for a concise description.

This algorithm takes as input a (truncated) series $\{S_k\}_{k=1}^{M}$ and produces as output $\{T^0_1=S_1, T^1_1, T^2_1, T^3_1,\ldots,  T^ {[[(M-1)/2]]}_1\}$ and $\{T^0_2=S_2,T^1_2,\ldots ,T^ { [[M/2]]-1 }_2\}$.  Here $[[R]]$ denotes the integer part of a real number $R$. We can summarize both parts into one series ${\cal W}\equiv\{S_1,S_2,T^1_1, T^1_2,T^2_1, T^2_2,  \ldots, \}$.

Depending on whether $M$ is odd  or even,  $T^ {[[(M-1)/2]]}_1$ respectively $T^ { [[M/2]]-1}_2$ should be adopted as best approximation to the limit value $S$.  Of course one should choose the parameter $M$ such that the difference between these two values is much smaller than the precision one is interested in. 

%


 Wynn's algorithm (\ref{Wynn1})  can lead to a division by very small numbers. This can lead to a termination of the numerical algorithm.  One can however implement simple stabilization procedures, which prevent a termination of the computation due to overflow caused by this issue \cite{WenigerReview}.

In this work we will apply Wynn's epsilon algorithm to compute the Regge path integrals and effective spin foam sums. To this end we have to choose several parameters, i.e. $C_{\rm min}$ and $C_{\rm step}$ in (\ref{Accsum}) and (\ref{Accint}) as well as the length of the series to which we apply Wynn's epsilon algorithms. The choice of these parameters influences the degree of convergence for the resulting series.  To control this rate of convergence, we define the maximal relative error
\ba\label{error}
{\cal E}= \text{max}_{k\in {\cal K}}  \frac{|{\cal W}_{k}- {\cal W}_{k-1}|}{\tfrac{1}{2}| {\cal W}_{k}+ {\cal W}_{k-1}|} \q ,
\ea
where ${\cal K}$ include the last 5 labels for the series $\{{\cal W}_k\}_k$, which results from Wynn's epsilon algorithm. For computations of the partition functions, we choose the above mentioned parameters so that ${\cal E}<10^{-6}$. (Often much smaller maximal errors could be reached, e.g. ${\cal E}<10^{-12}$.) For the computation of expectation values, 
we encountered a number of examples were it was hard to find parameters which led to such small relative errors.
The underlying reason is that the sums for the computation of the expectation values can become very small, see the example in Fig.~\ref{FigAcc2}.
 We therefore allowed relative errors up to ${\cal E}<10^{-3}$.

\subsection{Applications}

The non-linear sequence transformations, such as the higher order Shanks transform, are {\emph not} guaranteed to lead to fast convergence for all series. But for a series which can be approximated by the form (\ref{S5}) one can expect fast convergence.

In general we did observe  for the examples we computed for this paper a good convergence. We believe that the main reason for this is that the asymptotic behaviour of the Regge action for large (time-like) bulk areas is quite simple, and in particular linear in the summation/integration variable.\footnote{We observed that an action which asymptotes to a quadratic growth in the summation variable, might be harder to treat. The reason are pseudo saddle points, which can  appear because of the interplay between the frequency of the oscillations in the amplitude and the discretization of the variable, which defines the density with which the amplitude is probed. See Fig.~\ref{FigShell9} for an example of a pseudo saddle point. With a quadratic growth of the action in the summation variable such pseudo saddle points will appear and can lead to the partial sums exhibiting a stair-case like behaviour, that is oscillating around one value for a certain range of the cut-off, but then suddenly changing to oscillations around another value if this cut-off crosses a threshold value corresponding to a pseudo-saddle point. Such a behaviour is not well approximated by series of the form (\ref{S5}), and the Wynn epsilon algorithm has to be expected to fail in capturing the infinite sum. \label{footnoteQ}}

In the case that the action exhibits a saddle point along the Lorentzian integration/summation contour, that is for our shell model examples, this asymptotic behaviour will set in for bulk areas $B_t$ which are only slightly larger than the critical value. In this case it is advisable to define the partial sums with cut-offs above this critical value. Here it was generally sufficient to apply Wynn's epsilon algorithm to a series of length 100 for the spin foam sum (that is implement a 50th order Shank transform), and to a series of length less than 20  for the Regge integral, both for the computation of the partition functions and expectation values. 

In the case that the action  does not exhibit saddle points along the Lorentzian integration/summation contour, that is for our ball model examples, one can also identify easily an asymptotic regime. The resulting sums and integrals tend to lead to  much smaller numbers  than in the Lorentzian regime. We therefore experienced more difficulties to obtain the needed convergence for the computation of the expectation values, in particular for the spin foam sum. For some examples we had to consider a series of length 500, to obtain  relative maximal errors, which are smaller than $10^{-3}$.

The Regge integrals can be computed either by applying Wynn's epsilon algorithm, or by using a deformation of the integration contour into the complex plane, see \cite{ADP}. Naturally, we find in both cases the same results. 

The effective spin foam sums cannot be computed via a deformation of the `summation contour'. We will comment on this point in more detail in Section \ref{ResultsBall}. We therefore have to rely entirely on the method of non-linear sequence transformations. We will  see that the spin foam results in the shell model (that is for examples with saddle points along the integration contour) reproduce very well the Regge integrals and the expectation values approximate closely the classical solutions. This validates the method of non-linear sequence transformation, and justifies a posteriori the association of (anti-) limit values to divergent series, which appear for the computation of expectation values.

We note that the first order Shanks transform agrees with Aitken’s delta-squared process \cite{Aitken}. The latter has been applied previously in spin foams  \cite{Dona2022,Dona2}. The aim there was  to provide bounds for the results of an amplitude evaluation involving a cut-off.  Numerical constraints allowed only a cut-off on the order of $10^1$ and  the spin foam evaluation in question led to a non-oscillating series, truncated to an order of $10^1$ terms, which appears to converge extremely quickly if compared to  Fig.~\ref{FigAcc1}.

\section{Results}\label{Results}

\subsection{The shell model -- examples with a Lorentzian saddle point}\label{ResultsShell}

Here we will consider the shell model and a range of examples which have one saddle point along Lorentzian data.

\begin{figure}[ht!]
\begin{picture}(500,150)
\put(0,2){ \includegraphics[scale=0.6]{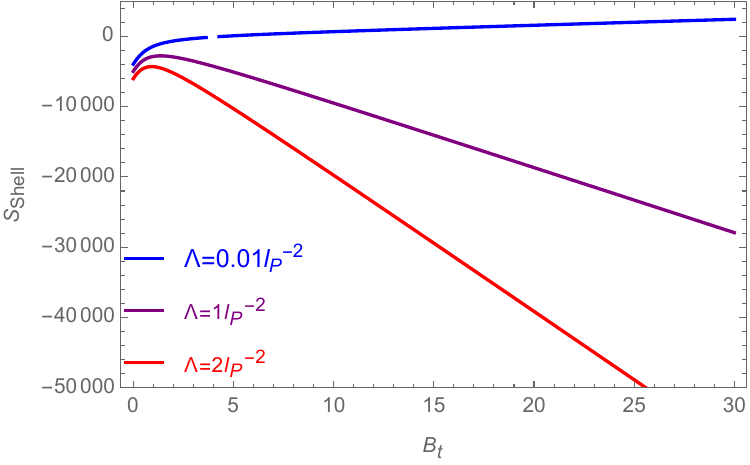} }
\put(260,2){ \includegraphics[scale=0.6]{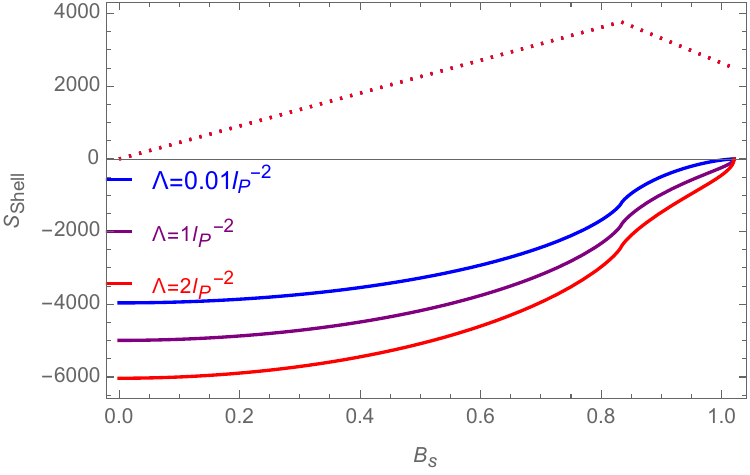} }
%
\end{picture}
\caption{ \footnotesize  The left and right panel show the action (in units of $\ell_P^2$) as function of the time-like bulk area $B_t$ (left panel) and of the space-like bulk area $B_s$ (right panel), both areas are given in terms of $\ell_P^2$.  Different colours for the graph encode different values for the cosmological constant.  We fixed as values for the boundary areas
$A_1=5\ell_P^2$ and $A_2=7.5 \ell_P^2$. 
For space-like areas the action has a real part (solid lines) and an imaginary part (dashed line). This imaginary part does not depend on the cosmological constant and is therefore the same for all shown cases.
The case with $\Lambda=0.01\ell_P^{-2}$ (in blue) does not have a Lorentzian critical point, but it has a Euclidean one. We will discuss examples with Euclidean critical points in the next section. \label{FigAction}}
\end{figure}

Regarding the numerical procedure, these examples with a saddle point along the original integration contour are typically easier to handle than the ball model examples where the saddle point is rather in the complex plane. One obtains fast convergence for e.g.  Wynn's epsilon procedure, if the minimal cut-off for the lapse variable is chosen such that it includes the saddle point. Beyond the saddle point the action does go over into a simple asymptotic behaviour, which is linear in our summation variable, see Fig.~\ref{FigAction}. 

\begin{figure}[ht!]
\begin{picture}(500,150)
\put(0,2){ \includegraphics[scale=0.6]{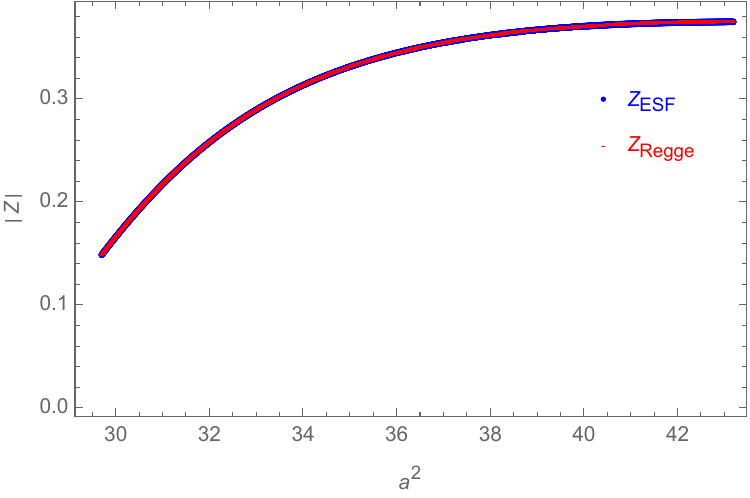} }
\put(260,2){ \includegraphics[scale=0.6]{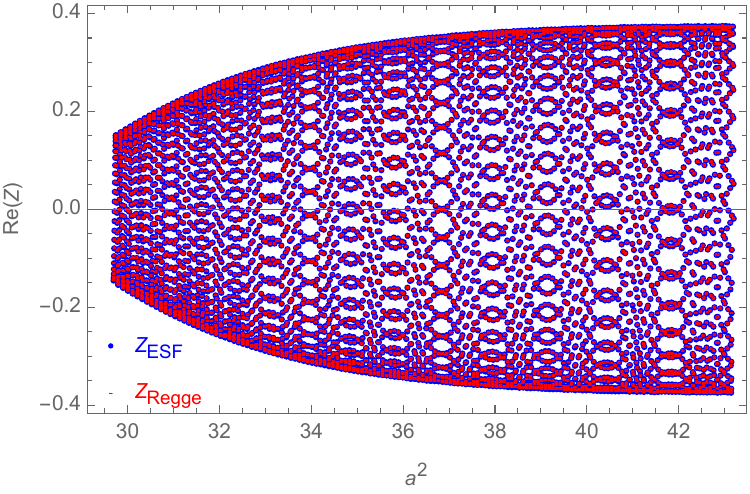} }
\put(33,130){$\Lambda=0.2\ell_P^{-2}$}
\put(295,130){$\Lambda=0.2\ell_P^{-2}$}

\end{picture}
\caption{ \footnotesize  The left and right panel show the absolute value and the real part of the spin foam sum and Regge integral, respectively.  Here we keep $A_1=5.5 \ell_P^2$ fixed and vary $A_2$ from $5.5\ell_P^2$ to $8\ell_P^2$. We translated this boundary area to a scale factor squared according to (\ref{translation}), and express it units of $\ell_P^2$. This applies also to all the following figures.  We  plotted only every fifths spectral value for the boundary area.  For both the absolute value and the real part differences between spin foam sum and Regge integral are not visible. \label{FigShell2}}
\end{figure}

\begin{figure}[ht!]
\begin{picture}(500,150)
\put(0,2){ \includegraphics[scale=0.6]{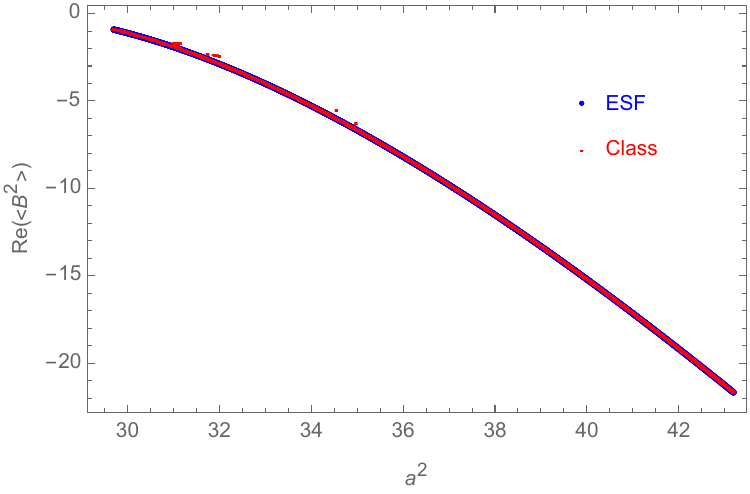} }
\put(260,2){ \includegraphics[scale=0.6]{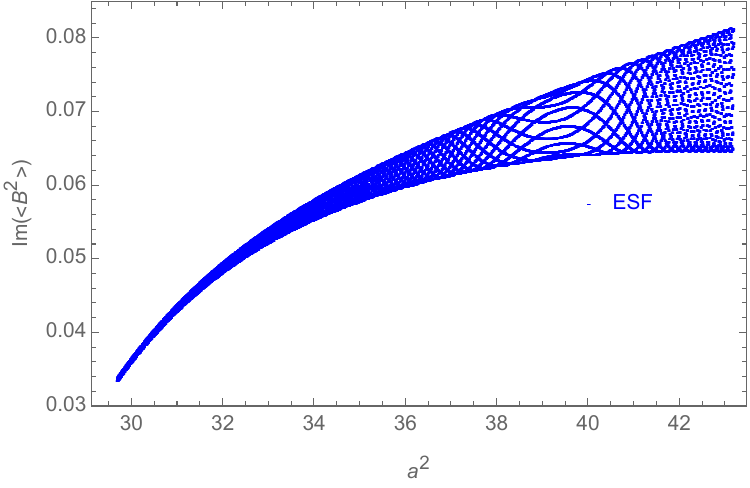} }
\put(33,110){$\Lambda=0.2\ell_P^{-2}$}
\put(295,110){$\Lambda=0.2\ell_P^{-2}$}

\end{picture}
\caption{ \footnotesize  The left panel shows the real part of the expectation value of the (signed) squared bulk area $B^2\equiv {\mathbb A}_{\rm blk}$ for the spin foam sum and the classical solution.  There are no visible differences. The right panel shows the imaginary part of the expectation value -- the imaginary part of the classical solution is vanishing.      Here we expressed $B^2$ in units of $\ell_P^4$ (also in the following figures). \label{FigShell3}}
\end{figure}

\begin{figure}[ht!]
\begin{picture}(500,150)
\put(0,2){ \includegraphics[scale=0.6]{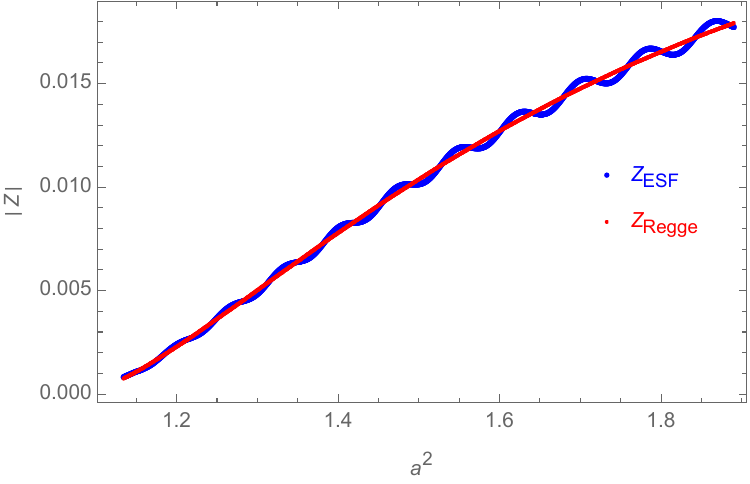} }
\put(260,2){ \includegraphics[scale=0.6]{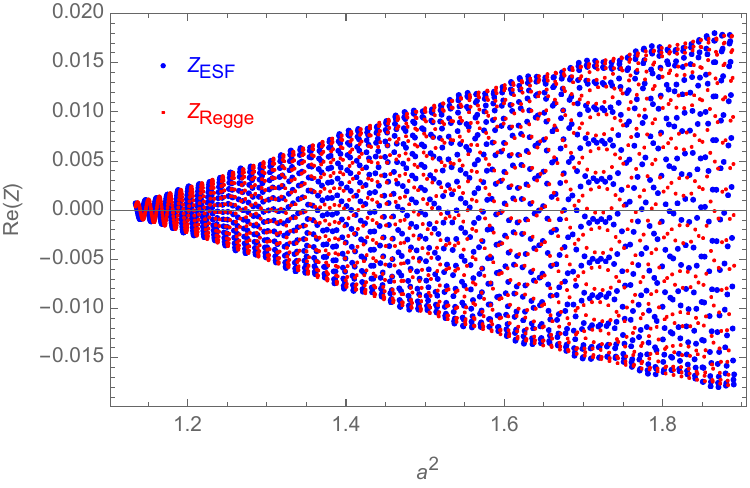} }
\put(36,131){$\Lambda=100\ell_P^{-2}$}
\put(296,131){$\Lambda=100\ell_P^{-2}$}
\end{picture}
\caption{ \footnotesize The left panel shows the absolute value of the spin foam sum and of the Regge path integral for the third example where $\Lambda=100\ell_P^2$. The right panel shows the real parts.  We now do see (small) differences between spin foam sum and Regge integral for both the absolute value and the real part. Here we again plot the partition functions for every spectral value in the shown range for the outer boundary area. \label{FigShell10}}
\end{figure}

The results will show:
\begin{itemize}
\item
The difference between including and excluding the hinge irregular region is relatively small. E.g. the absolute values of the differences in the spin foam sum for the first example below are of the order $10^{-8}$, for the second example $10^{-6}$  and for the third example $10^{-5}$. We will show here the version where we include the hinge irregular region into the path integral. There would not be a visible difference in the plots for the alternative choice.
\item
The difference between Regge path integrals and effective spin foam sums are relatively small, e.g. see Fig.~\ref{FigShell2}. 
\item
The  expectation value for the bulk area squared is well approximated by its classical value, e.g. see Fig.~\ref{FigShell3}.  
\item 
Larger deviations can be found for large values of the cosmological constant, see Fig.~\ref{FigShell10} and Fig.~\ref{FigShell11}.
\end{itemize}

We will illustrate these features with a few examples below.

Let us start with our first example. Here we choose $\Lambda=0.2 \ell_P^{-2} $  and fix as the smaller boundary area $A_{1}:=\sqrt{ \mathbb{A}_{\rm bdry_1}}=5 \ell_P^2$. For the larger boundary area we consider a range from
$A_{2}=  5.5\ell_P^2$ to $A_{2}=8\ell_P^2$.  We can translate these boundary areas to the scale factor squared: $a^2_1\approx 27  \ell_P^2$ and $a^2_2 \approx 29.7 \ell_P^2, \ldots, 43.2 \ell_P^2$. To do so we equate the three-volume of the hypersurface of the shell model with the three-volume of the hypersurface in the mini-superspace model, see \cite{DGS}. Note that in the mini-superspace model, there are Lorentzian solutions for the lapse function, if both boundary values for the scale factor squared are larger than $a^2_\Lambda=15 \ell_P^2$.

The main difference between the Regge path integral and the effective spin foam sum is that in the former case we integrate over the bulk variable and in the second case we rather sum over the spectral values \eqref{BulkSp}.  We expect to see large differences between the results if, around the classical solution, there are only very few spectral values per oscillation of the amplitude. 

\begin{figure}[ht!]
\begin{picture}(500,150)
\put(0,2){ \includegraphics[scale=0.6]{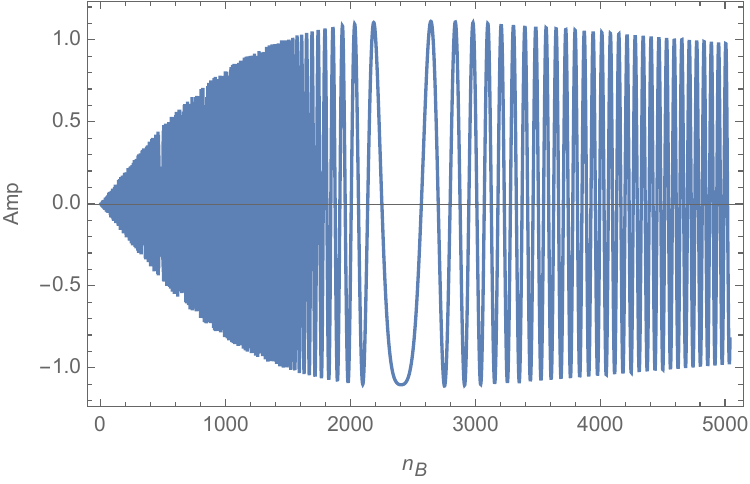} }
\put(260,2){ \includegraphics[scale=0.6]{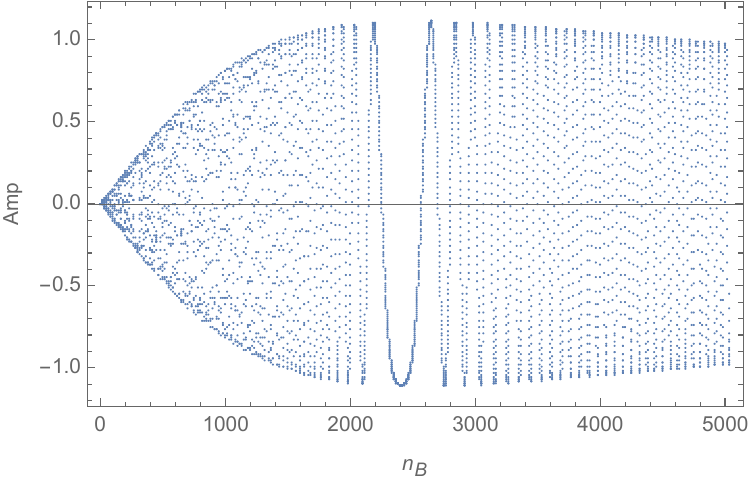} }
\put(33,130){$\Lambda=0.2\ell_P^{-2}$}
\put(295,130){$\Lambda=0.2\ell_P^{-2}$}
%
\end{picture}
\caption{ \footnotesize These plots show the amplitude (including the measure term) as a function of the (time-like) summation parameter $n_B$ for the example with $\Lambda=0.2\ell_P^{-2}$.  (That is $B_t=\ell_P^2 n_B/n_e$ with $n_e=720$.) On the left panel we use a continuous plot, on the right panel we show only the amplitude values for the discrete values of $n_B$, which we sum over.  Note that if we would have used a spectrum $B_t=\ell_P^2 n_B$, the $5040=7ne$ points shown in the right panel would reduce to just 7 points. \label{FigShell1}}
\end{figure}

Fig.~\ref{FigShell1} shows that this is not the case, and that we can determine very well the saddle point also with the restriction of the bulk variable to its spectral values \eqref{BulkSp}.  We indeed find that the Regge path integral and the effective spin foam sum approximate each other very well see Fig.~\ref{FigShell2}.

\begin{figure}[ht!]
\begin{picture}(500,150)
\put(0,2){ \includegraphics[scale=0.6]{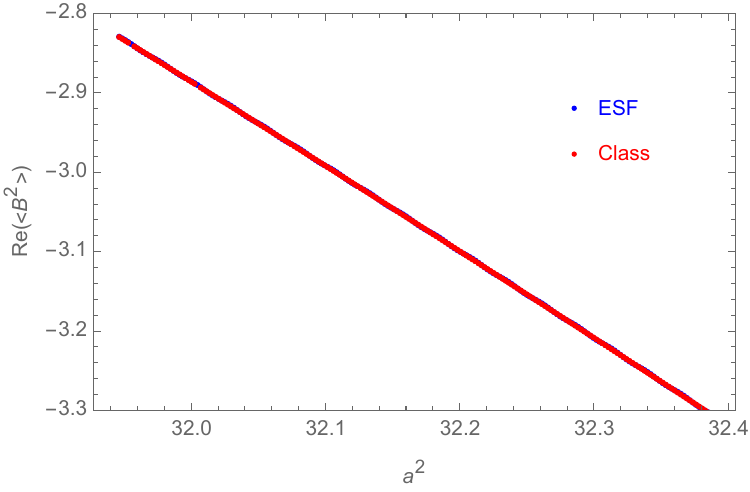} }
\put(260,2){ \includegraphics[scale=0.6]{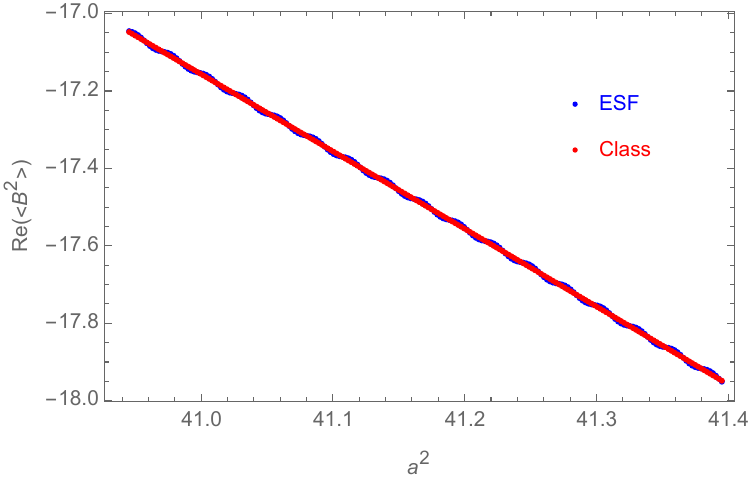} }
\put(33,30){$\Lambda=0.2\ell_P^{-2}$}
\put(295,30){$\Lambda=0.2\ell_P^{-2}$}

\end{picture}
\caption{ \footnotesize Here we zoom into a smaller region of the plot shown on the left panel in Fig.~\ref{FigShell3}. 
The left panel shows the expectation value for smaller values of the scale factor squares for the outer boundary $a^2_2$ (or smaller differences $(a_2-a_1)^2$) and the right panel for larger differences.  In the latter case we do observe very small oscillations of the expectation value around the classical value.
 \label{FigShell4}}
\end{figure}

Fig.~\ref{FigShell3} compares the (effective spin foam) expectation value of the bulk area squared with the classical solution. The differences for the real part of this expectation values are however not visible in this plot.  Tiny oscillations of the expectation value are only visible for larger values of the scale factor, if one zooms in, see Fig.~\ref{FigShell4}.  In Fig.~\ref{FigShell3}  we also show the imaginary part of the expectation value. The fact that is does not vanish can be understood as a quantum effect, the ratio of imaginary and real part is however quite small, namely $\approx 10^{-3}$.

\begin{figure}[ht!]
\begin{picture}(500,150)
\put(0,2){ \includegraphics[scale=0.6]{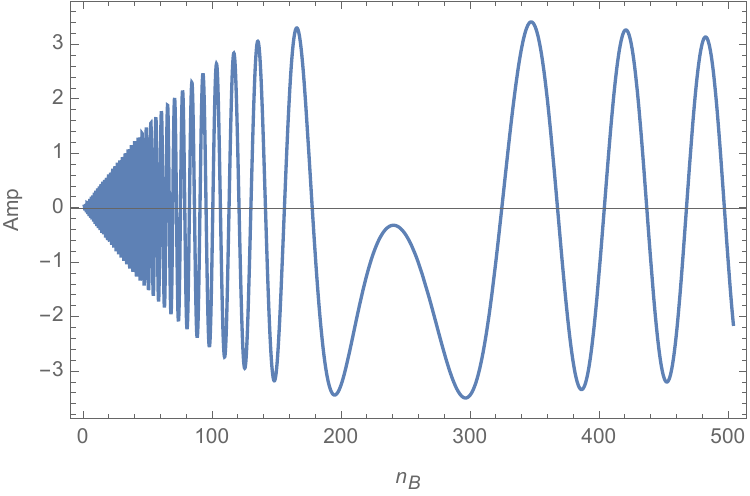} }
\put(260,2){ \includegraphics[scale=0.6]{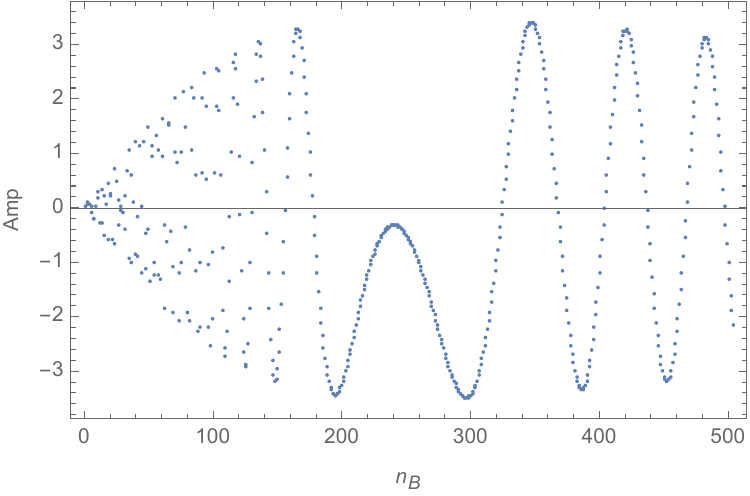} }
\put(33,130){$\Lambda=2\ell_P^{-2}$}
\put(293,130){$\Lambda=2\ell_P^{-2}$}
\end{picture}
\caption{ \footnotesize 
These plots show the amplitude (including the measure term) as a function of the (time-like) summation parameter $n_B$ for the example with $\Lambda=2\ell_P^{-2}$. This example is connected by scaling (\ref{scalingAct}) to the example shown in Fig.~\ref{FigShell1}. Due to this scaling one has approximately the same number of spectral points per oscillation for the two examples, and could therefore expect the same quality of approximation of the Regge integral by the spin foam sum. \label{FigShell5}}
\end{figure}

For the second example we choose $\Lambda=2 \ell_P^{-2}$. Note that this value is $10$ times the value of the cosmological constant in the first example. We choose the boundary areas (or squared scale factors) to be $1/10$ the values in the first example. The actions in the first and second example are then connected by the following scaling behaviour
\ba\label{scalingAct}
S_{\rm Shell}\left(\frac{1}{c} A_1, \frac{1}{c}  A_2,   \frac{1}{c}  B  ; \, c\,\Lambda\right) \,=\,  \frac{1}{c} \,S_{\rm Shell} (A_1,  A_2,  B  ; \,\Lambda) \q .
\ea
This scaling behaviour lets us expect the same number of (bulk area) spectral values per oscillation of the amplitude for examples that are connected by scaling.  Indeed Fig.~\ref{FigShell5} shows that the saddle point is still very well recognizable in this second example, although we show only $1/10$ of the spectral values for the bulk area as compared to the first example.

\begin{figure}[ht!]
\begin{picture}(500,150)
\put(0,2){ \includegraphics[scale=0.6]{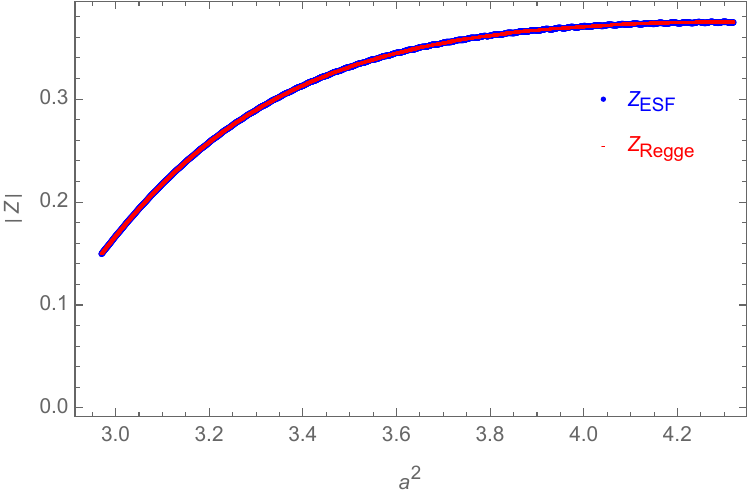} }
\put(260,2){ \includegraphics[scale=0.6]{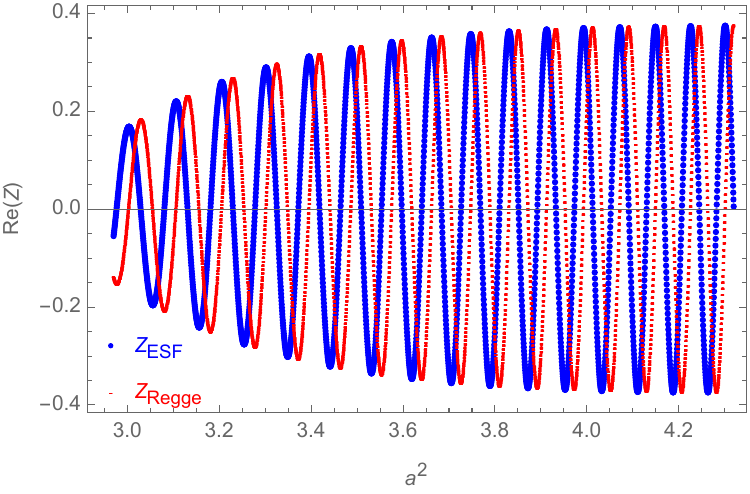} }
\put(33,130){$\Lambda=2\ell_P^{-2}$}
\put(293,130){$\Lambda=2\ell_P^{-2}$}
\end{picture}
\caption{ \footnotesize  The left panel shows the absolute value of the spin foam sum and of the Regge path integral for the second example. The right panel shows the real parts. In the latter we see a small shift in the phase between spin foam sum and Regge path integral.  Here we plot the partition functions for all spectral values in the shown range for the outer boundary area. \label{FigShell6}}
\end{figure}

Fig.~\ref{FigShell6} compares the absolute value of the Regge and effective spin foam partition function, and we again see that these approximate each other very well. The real parts of the partition functions differ by a small shift.

\begin{figure}[ht!]
\begin{picture}(500,150)
\put(0,2){ \includegraphics[scale=0.6]{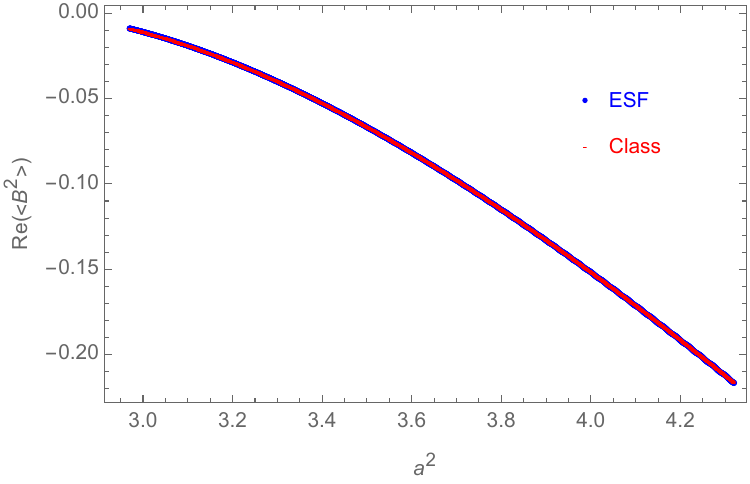} }
\put(260,2){ \includegraphics[scale=0.6]{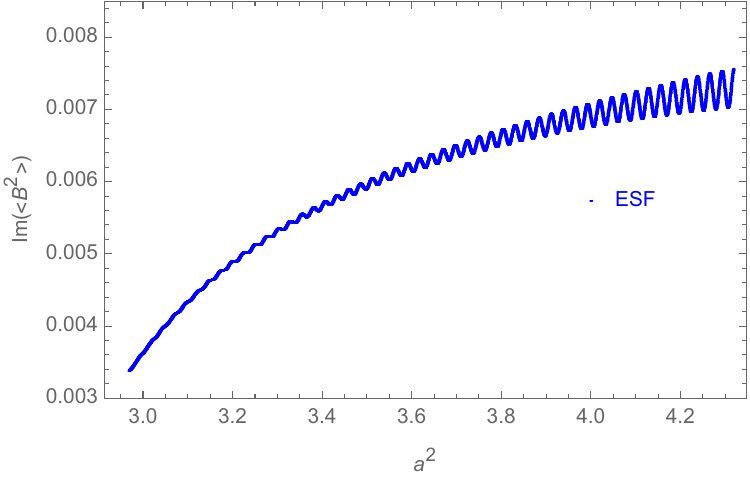} }
\put(37,100){$\Lambda=2\ell_P^{-2}$}
\put(297,100){$\Lambda=2\ell_P^{-2}$}
\end{picture}
\caption{ \footnotesize  The left panel shows the real part of the expectation value of the (signed) squared bulk area  for the spin foam sum and the classical solution, for the second example.  There are no visible differences. The right panel shows the imaginary part of the expectation value, whose size is approximately  three percent of the size of the real part. \label{FigShell7}}
\end{figure}

\begin{figure}[ht!]
\begin{picture}(500,150)
\put(0,2){ \includegraphics[scale=0.6]{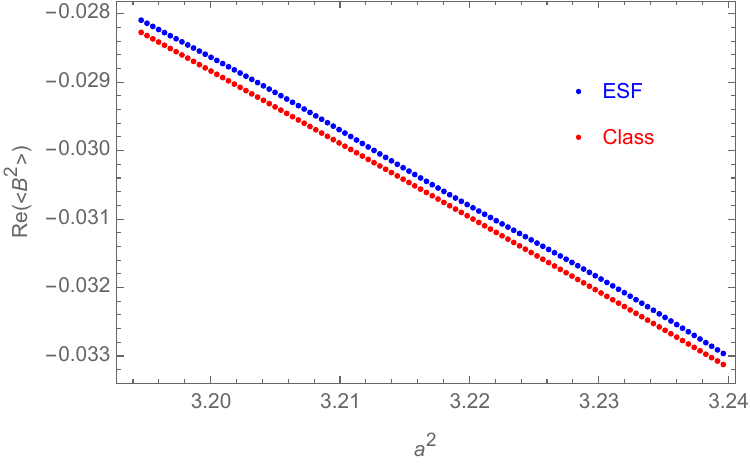} }
\put(260,2){ \includegraphics[scale=0.6]{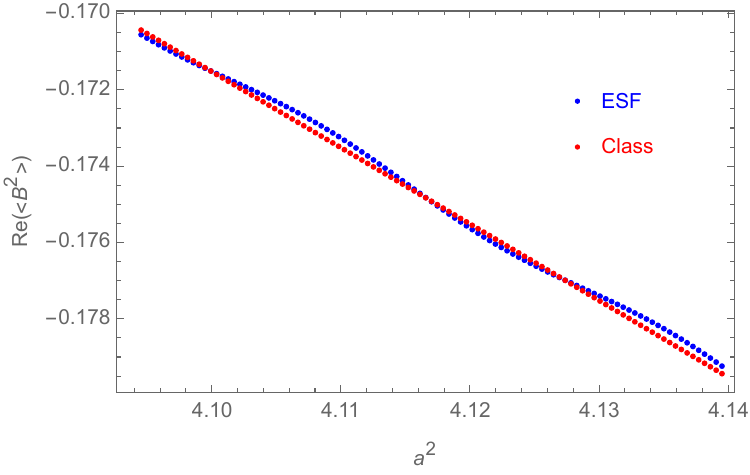} }
\put(40,90){$\Lambda=2\ell_P^{-2}$}
\put(300,90){$\Lambda=2\ell_P^{-2}$}
\end{picture}
\caption{ \footnotesize  Here we zoom into the plot shown on the left panel in Fig.~\ref{FigShell7}. We see a small  shift for smaller values for the outer squared scale factor and small oscillations for larger values for the outer squared scale factor. \label{FigShell8}}
\end{figure}

We depict the expectation values in Fig~\ref{FigShell7} and Fig~\ref{FigShell8}.  As expected, the real part of the expectation values are connected by the scaling behaviour (\ref{scalingAct}). This extends to the size of the imaginary part. 

However, zooming in, see Fig~\ref{FigShell8}, we find that the differences between the expectation value and the classical value appear (in their relative size) to be somewhat larger than in the first example, shown in Fig~\ref{FigShell4}.

We choose the third and last example to showcase an example with larger differences between the Regge and effective spin foam path integral. We found that these differences are enhanced for large values of the cosmological constant, and therefore choose $\Lambda=100\ell_P^2$.  The boundary values are chosen as $A_1=0.2 \ell_P^2$ ($a_1^2\approx 1.1\ell_P^2$) and $A_2=0.22\ell_P^2, \ldots, 0.4\ell_P^2$ ($a_1=1.2\ell_P^2,\ldots, 2.1 $). 

\begin{figure}[ht!]
\begin{picture}(500,150)
\put(0,2){ \includegraphics[scale=0.6]{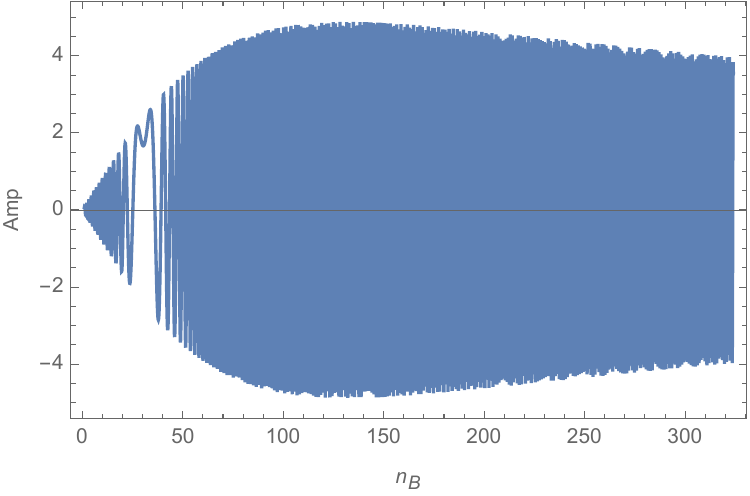} }
\put(260,2){ \includegraphics[scale=0.6]{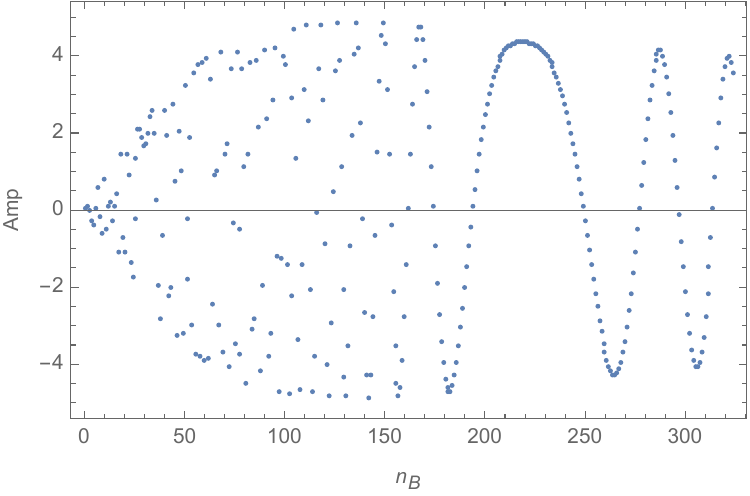} }
\put(31,132){$\Lambda=100\ell_P^{-2}$}
\put(291,132){$\Lambda=100\ell_P^{-2}$}
\end{picture}
\caption{ \footnotesize These plots show the amplitude as a function of the (time-like) summation parameter $n_B$ for the example with $\Lambda=100\ell_P^{-2}$. On the left panel we show the amplitude as a function of a continuous parameter $n_B$. Here we see a (true) saddle point around $n_B=30$. On the right we show the amplitude as a function of the discrete summation parameter $n_B \in {\mathbb N}$.  Due to the size of the spectral gap we do not see anymore the (true) saddle point. One rather has a pseudo saddle point around $n_B=220$, which arises trough interference between the discretization and the frequency of the oscillations in the amplitude. \label{FigShell9}}
\end{figure}

\begin{figure}[ht!]
\begin{picture}(500,150)
\put(0,2){ \includegraphics[scale=0.6]{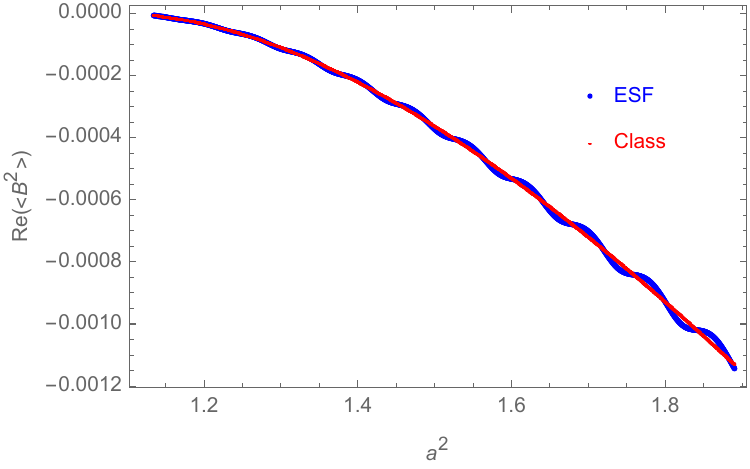} }
\put(260,2){ \includegraphics[scale=0.6]{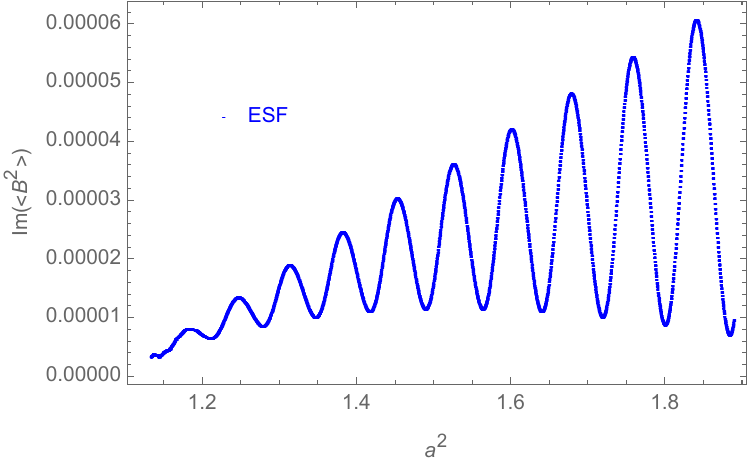} }
\put(42,113){$\Lambda=100\ell_P^{-2}$}
\put(302,113){$\Lambda=100\ell_P^{-2}$}
\end{picture}
\caption{ \footnotesize Here we show real and imaginary part of the expectation value for the (signed) bulk area squared and compare with the classical solution. The expectation value still approximates the classical value very well, despite the fact that the true saddle point is washed out by the discretization (see Fig.~\ref{FigShell9}). We see however small deviations with oscillatory behaviour. \label{FigShell11}}
\end{figure}

Fig.~\ref{FigShell9} shows the amplitude as a function of the continoues bulk variable and as a function of the discrete bulk variable.  There is a classical saddle point at $j_B \approx 30$, this classical saddle point is however not visible in the discrete plot. We see however a so-called pseudo-saddle point at $j_B \approx 225$: such saddle points appear if the frequency of the spectral values is similar to the frequency of the oscillations. 

One could therefore expect rather large differences between the Regge path integral and the effective spin foam path integral, and also large deviations of the effective spin foam results from the classical limit. We will see however that the classical results are still reproduced to a surprising degree.

Fig.~\ref{FigShell10} compares the absolute value of the Regge and effective spin foam partition functions. Compared to the previous examples there are visible differences, which are however relatively small. The deviations increase for growing boundary scale factor $a^2_2$. This could be due to the fact that there is a pseudo saddle point for larger values of this boundary scale factor, but we could identify no such pseudo saddle point for smaller values of $a_2^2$.

\begin{figure}[ht!]
\begin{picture}(500,150)
\put(0,2){ \includegraphics[scale=0.6]{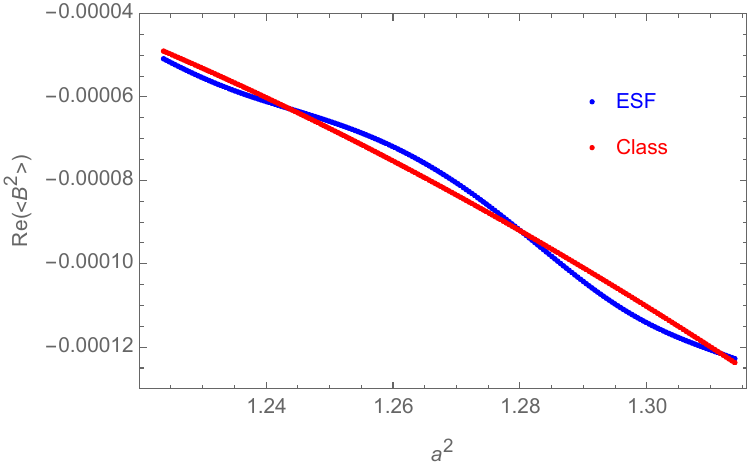} }
\put(260,2){ \includegraphics[scale=0.6]{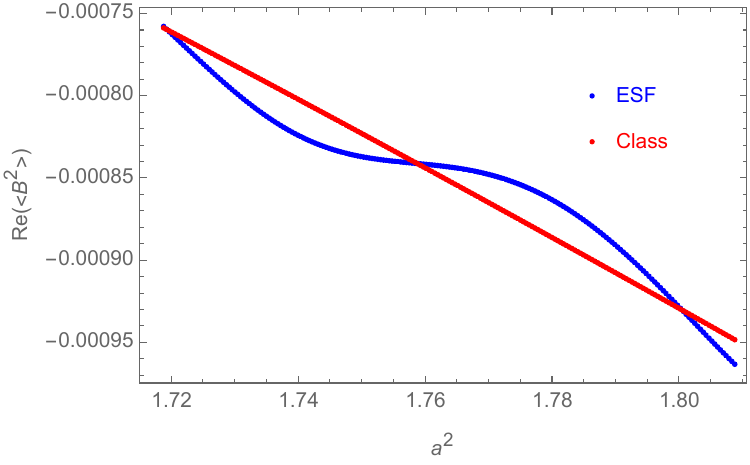} }
\put(45,30){$\Lambda=100\ell_P^{-2}$}
\put(305,30){$\Lambda=100\ell_P^{-2}$}
\end{picture}
\caption{ \footnotesize Here we zoom into the plot shown on the left panel in Fig.~\ref{FigShell11}. We now see visible differences between the expectation value and the classical value both for small and larger value of the outer scale factor. \label{FigShell12}}
\end{figure}

We compare in Fig.~\ref{FigShell11}  the effective spin foam expectation value for the squared bulk area with the classical value. The real part of the expectation value does approximate the classical value quite well, given that we cannot identify the classical saddle point in the right panel of Fig.~\ref{FigShell9}. We notice however also small oscillations of the expectation value around the classical value, which become more noticable when we zoom into a smaller range for the boundary scale factor, see Fig.~\ref{FigShell12}. The oscillations again increase, if we increase the values for $a^2_2$.

Overall we find that for the shell model, with boundary values chosen such that we have classical saddle points for negative values of the bulk area square (and thus satisfying the Lorentzian triangle inequalities), we have relatively small differences between the Regge path integral and the effective spin foam sum on the one hand, and the effective spin foam sum expectation value and the classical solution on the other hand. 

This does change drastically for examples where we have only a saddle point for a (positive) value of the bulk area square, which satisfY the Euclidean triangle inequalities. Such saddle points can be interpreted to have imaginary lapse and to describe a Euclidean geometry.

\subsection{The ball model -- examples with a Euclidean saddle point}\label{ResultsBall}

Here we will consider the ball model, which implements a boundary value $A_1=0$ and thus $a_1^2=0$. We will choose the second boundary value sufficiently small so that the classical action has a  saddle point for a positive bulk area, so that Euclidean triangle inequalities are satisfied.

 The path integral, as a function of the second boundary value, does the so-called no-boundary wave function \cite{HartleHawking}.  In the continuum mini-superspace model the path integral can be reduced\footnote{ The path integral includes a priori integrals over the scale factors at each time step and over lapse at each time step. After a variable transformation, the action can be made to be quadratic in a variable related to the scale factors, and this variable can be integrated out. The lapse variables are gauge fixed leaving only a global lapse parameter to be integrated over.} to a one-dimensional integral over a global lapse parameter. Starting from a Lorentzian contour one can deform the contour to go along Euclidean data, and apply a saddle point approximation \cite{TurokEtAl} or evaluate this integral numerically \cite{ADP}. The work \cite{ADP} considered the Regge path integral based on the ball model which approximates continuum time evolution with one time step. This reduces the `path integral' to an integral over one variable, which can be interpreted as lapse. This integral was evaluated using a deformation of the contour to Euclidean data, along which the integral is exponentially decaying. This allowed a numerical integration along this contour. Despite the rather coarse approximation of the continuous time evolution with one simplicial time step, the Regge model provided an astonishingly accurate approximation of the continuum mini-superspace results \cite{ADP}.
 
 Here we will use instead of a contour deformation a direct evaluation of the Lorentzian path integral, based on using acceleration operators for series convergence. For the Regge integral we find exactly the same results as with the contour deformation.
 
 For (effective) spin foams, we replace the integral with a discrete sum and we thus cannot apply a contour deformation.  We will thus rely on the acceleration operator technique.  In contrast to the cases with Lorentzian saddle points,  we will find for the cases with Euclidean saddle points that the effective spin foam sum substantially deviates from the Regge integral. This deviation grows with the difference between the initial and final scale factors, but appears already for quite small differences. On the other hand the deviation does decrease, if we refine the spectrum for the bulk variable. As one can expect, the deviations vanish when we consider the infinite refinement limit.

We will thus find a drastic difference between the regime with Lorentzian and Euclidean saddle points. In the cases with Lorentzian saddle points, we found only small differences between the Regge and effective spin foam models, and reproduced the classical expectation value even in cases where the classical saddle points could not be identified in the discrete amplitude plot.  That is, the Lorentzian regime is not very sensitive to a discretization of the bulk variable.  In contrast the Euclidean regime appears to be sensitive to such a discretization. 

A possible explanation for this sensitivity might be rooted in the fact that in the Lorentzian regime the saddle points are on the initial integration contour, whereas in the Euclidean regime there are no saddle points on the initial integration contour. 

If we perform the Regge integral for the examples in the Euclidean regime along the initial contour, the result relies on  subtle cancellations between different parts of the amplitude, occurring across the entire integration range. This leads to an overall exponential suppression, which (for larger differences between initial and final scale factors) can lead to very small absolute values for the result of the integral, e.g. on the order of $10^{-6}$. These subtle cancellations are easily disturbed by a discretization of the bulk variable. Indeed we will see that, whereas cancellations still occur in the discrete case, the suppression effect works only to a certain degree, and one will not reach such small numbers as for the continuous integral. 

On the other hand, if we do a deformation of the integration contour to Euclidean data, we rely on an analytical continuation. The concept of harmonic functions and analytical continuation can be extended to functions of discrete variables \cite{DiscreteHarm}. Such discrete analytical continuations behave however quite differently form their continuum counterparts \cite{Unpublished}. E.g. whereas the analytical continuation of $\exp(\imath x)$ into the upper complex half plane lead to a decaying function, this is generally\footnote{The discrete analytical continuation can be uniquely specified if one provides the function values for positive real integers and positive imaginary integers. Trying different choices for the values of the function for positive imaginary integers, we always found directions in the positive complex half-plane in which the discrete analytical continuation diverges.} not the case for the discrete version of this function. 

In view of this discussion we conjecture that integrals with no saddle points along the initial contour are quite sensitive to a discretization of the integration variable. That is,  phenomena like tunnelling can be affected. Interestingly, we find that the probability for tunnelling tends to increase.

Although we provide a general reason for the deviation between the Regge and effective spin foam model in the Euclidean regime, we  want to point out two caveats that will be addressed in future work: the first one is that the results depend on the density of the spectrum of the integration variable. In Section \ref{SecSpectra} we provided a heuristic reasoning for our particular choice in this simplicial mini-superspace model.  It would be highly desirable to derive such a spectrum more directly from the non-symmetry reduced theory. This task can be compared with the still open question to derive loop quantum cosmology more properly from the full theory \cite{LQCfromFull}. The second caveat is that we consider only one time step in our simplicial model. As we explained above, we find that the one-time-step Regge  integral approximates the continuum result to an astonishing degree, see \cite{ADP}.  For the effective spin foam model we find however that the deviations from the Regge integral grow with the difference between the boundary values. This could possibly change if we consider multiple time steps, and ultimately the continuum limit.

\begin{figure}[ht!]
\begin{picture}(500,150)
\put(0,2){ \includegraphics[scale=0.6]{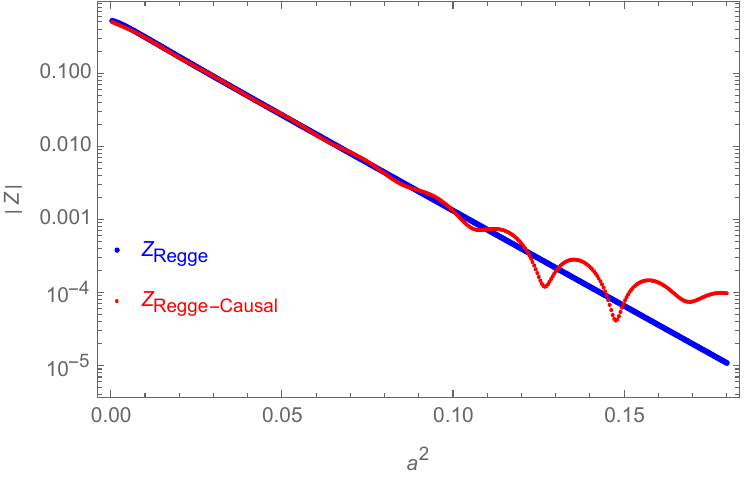} }
\put(250,2){ \includegraphics[scale=0.6]{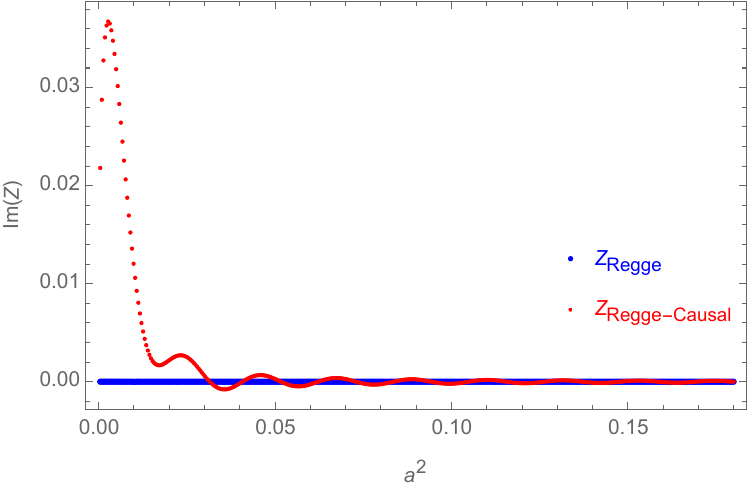} }
\put(65,130){$\Lambda=0.2\ell_P^{-2}$}
\put(325,130){$\Lambda=0.2\ell_P^{-2}$}
\end{picture}
\caption{\footnotesize  Here we compare the absolute value (left panel, with a logarithmic plot) and imaginary part (right panel) of the Regge integral for two different choices: For $Z_{\rm Regge}$ we integrate over both the hinge regular and irregular configurations, for $Z_{\rm Regge-Causal}$ we only integrate over the hinge regular configurations. Including the irregular configurations into the integral leads to a continued exponential decay of the partition function, which is also real. These two features are shared with the path integral for the continuum mini-superspace model. \label{Fig-Causal}}
\end{figure}
 
We will now present the numerical results in more detail. To begin with we consider the difference between the Regge integral with and without including the region with irregular light cone structure. Fig. \ref{Fig-Causal} compares these two integrals for the case $\Lambda=0.2 \ell_P^{-2}$.  We see that the choice where we do include the irregular region leads to a real result for the path integral, whereas not including the irregular region leads to a non-vanishing imaginary part. The continuum mini-superspace model (which does not feature such an irregular region) does lead to a real result. Furthermore, the path integral, which includes the irregular region, shows an exponential decay as a function of the squared scale factor, similar to the continuum mini-superspace model. In contrast, the choice which excludes the irregular region, shows an exponential decay only up to a certain threshold value. 

Thus, if we wish to have a close approximation of the continuum mini-superspace results we should include the irregular region. In the following we will only show the choice where we integrate or sum over the irregular region.

For the first example we will choose $\Lambda=0.2 \ell_P^2$ and $\gamma=0.1$. For the ball model the initial boundary area or initial scale factor (squared) is always vanishing. For the final area\footnote{In the plots we again translate the boundary areas to scale factors squared by comparing the 3-volumes in the simplicial model with the 3-volume of the spatial hypersurfaces in the mini-superspace model, see \cite{DGS}.} we will choose a range from $A=1\times(\gamma /n_t)\ell_P^2  $, that is the smallest non-vanish value, to $A = 453\times (\gamma /n_t)\ell_P^2$. Note that this is only a small part  (namely around 1 percent) of the range for the boundary area $A$ that leads to Euclidean saddle points and approximates well the continuum behaviour \cite{DGS}. This regime allows values up to $A = 45300\times (\gamma /n_t)\ell_P^2$. But we will find significant deviations from the Regge path integral already for this small regime.

\begin{figure}[ht!]
\begin{picture}(500,150)
\put(0,2){ \includegraphics[scale=0.6]{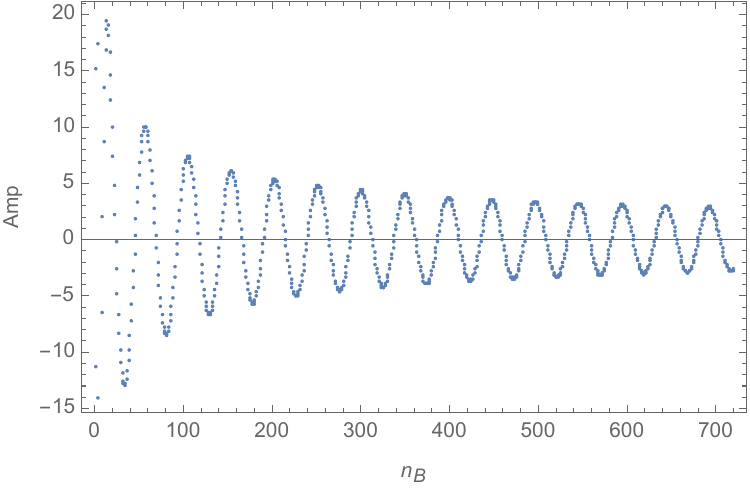} }
\put(250,2){ \includegraphics[scale=0.6]{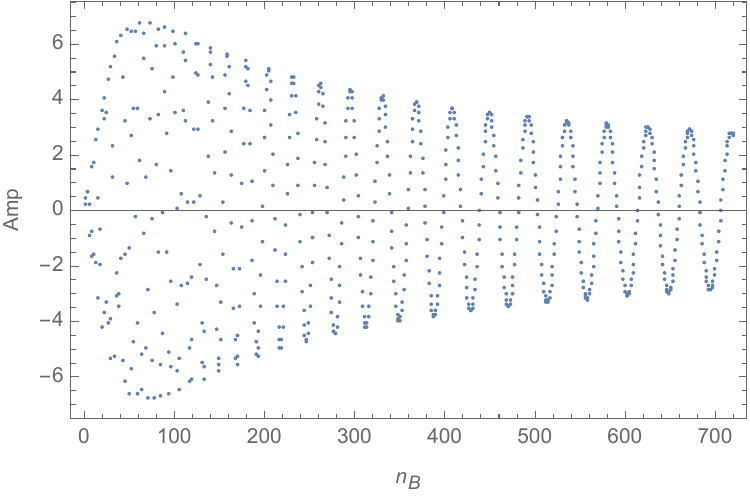} }
\put(75,130){$\Lambda=0.2\ell_P^{-2},\, A=\frac{\gamma}{n_t}\ell_P^2$}
\put(335,130){$\Lambda=0.2\ell_P^{-2}, \, A=300\frac{\gamma}{n_t}\ell_P^2$}
\end{picture}
\caption{\footnotesize  Here we show the amplitude for a smaller value (left panel) and a larger value (right panel) of the boundary area $A$ as function of the discrete (time like) summation variable $n_B$. We see that for larger boundary areas (equivalent to having larger differences between the outer and inner boundary, as the area for the inner boundary is vanishing) the discretization does not capture fully the oscillations of the amplitude over a larger range of the summation variable $n_B$. One can therefore expect that the difference between Regge integral and spin foam sum is larger for larger outer areas. \label{FigBFreq}}
\end{figure}

We will find larger deviations for larger values of the outer area. This can be explained by comparing the spectral gap for the bulk variable with the frequency of the oscillations of the amplitude. Fig.~\ref{FigBFreq} shows that this ratio is smaller for larger outer areas. We  therefore expect larger deviations in this case.

\begin{figure}[ht!]
\begin{picture}(500,150)
\put(0,2){ \includegraphics[scale=0.6]{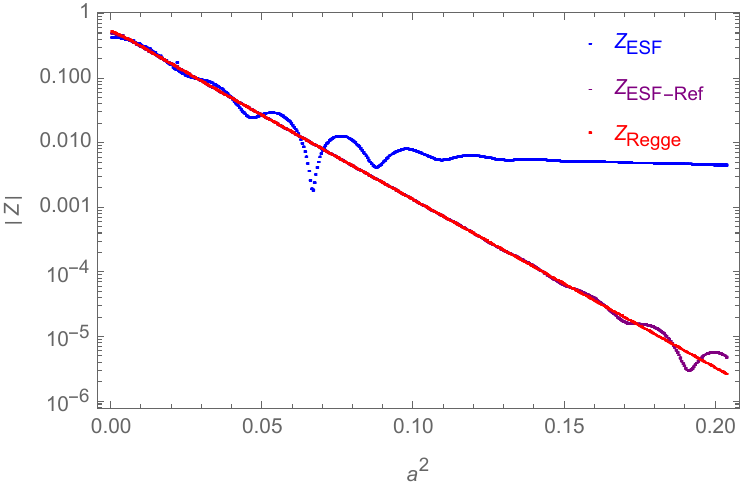} }
\put(250,2){ \includegraphics[scale=0.6]{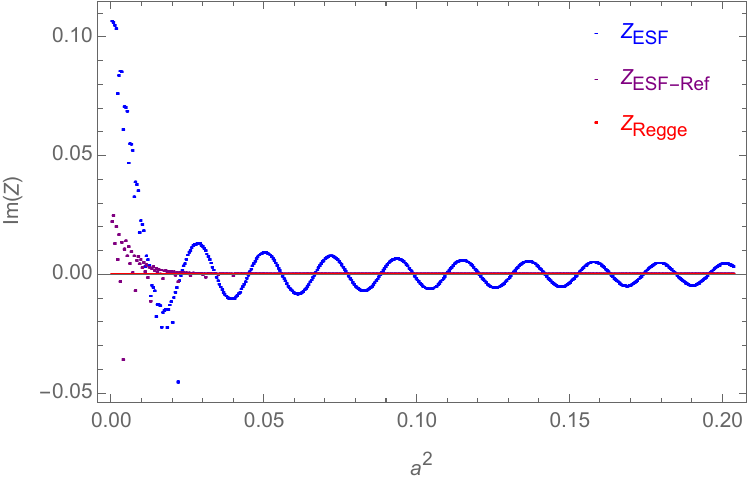} }
\put(75,130){$\Lambda=0.2\ell_P^{-2}$}
\put(335,130){$\Lambda=0.2\ell_P^{-2}$}
\end{picture}
\caption{\footnotesize  Here we show the absolute value (with a logarithmic plot on the left panel) and the imaginary part (right panel) of the partition functions for the ball model. We compare the Regge integral (red) with the effective spin foam sum (blue), with spectral values as described in (\ref{BulkSp}), and the effective spin foam sum (purple), where we refine the spectrum by a factor of 10.  We see that the effective spin foam sum does deviate from the Regge path integral for quite small values of the squared scale factor. That is the discretization (via the discrete spectra) of the integration variable does interfere with the subtle mechanism that leads to destructive interference and an exponential decay of the partition function. Refining the spectrum increases the range for the scale factor, in which the spin foam sum does approximate the Regge integral reasonable well. The imaginary part of the Regge integral does vanish, for the spin foam sums one does find a non-vanishing imaginary part. \label{FigB1}}
\end{figure}

This is confirmed by comparing the results for the effective spin foam sum with the Regge integral in Fig.~\ref{FigB1}. The (logarithmic) plot for the absolute value of the spin foam sum shows that it deviates from the Regge integral, starting with  relative small values for the boundary area or, equivalently the squared boundary scale factor. Refining the spectral values for the bulk variable by a factor of 10 one can increase the range where the spin foam sum and Regge integral agree.

\begin{figure}[ht!]
\begin{picture}(500,150)
\put(125,2){ \includegraphics[scale=0.6]{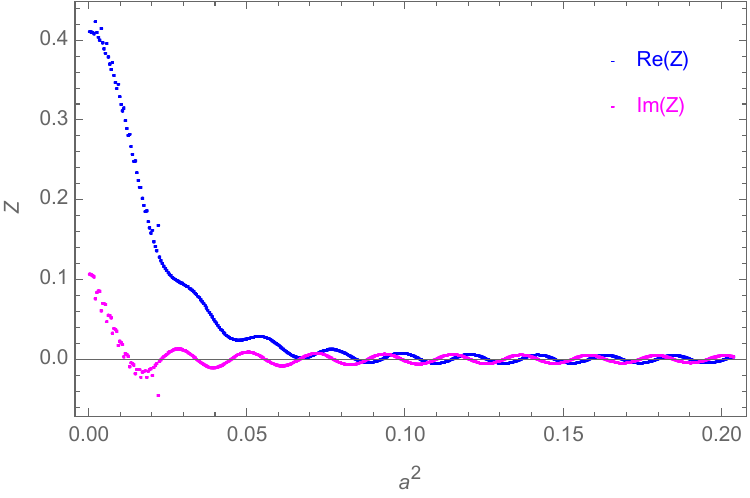} }
\put(170,130){$\Lambda=0.2\ell_P^{-2}$}
\end{picture}
\caption{\footnotesize  Here we show real and imaginary part of the spin foam partition function. The oscillations of the imaginary part are of the same magnitude as the oscillations of the real part, starting with quite small values of the scale factor (squared). \label{FigB1b}}
\end{figure}

Correspondingly, we find also a non-vanishing imaginary value for the spin foam sum, whereas it does vanish for the Regge integral. The value for the imaginary parts are larger for very small values of the boundary area. But comparing the size of the imaginary part with the absolute value we see that the relative size of the imaginary part grows for larger boundary areas and that the amplitude becomes fully oscillatory, see Fig.~\ref{FigB1b}.

\begin{figure}[ht!]
\begin{picture}(500,150)
\put(0,2){ \includegraphics[scale=0.6]{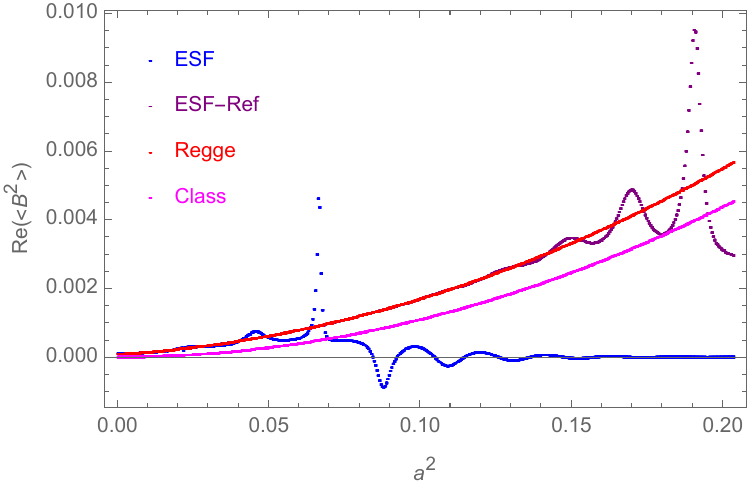} }
\put(250,2){ \includegraphics[scale=0.6]{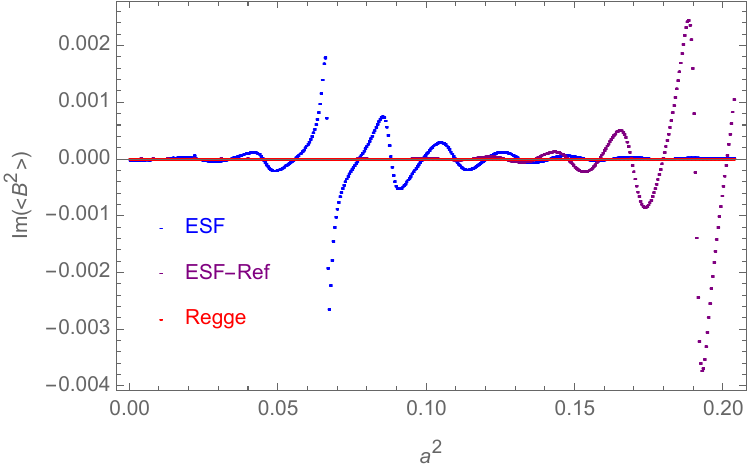} }
\put(95,125){$\Lambda=0.2\ell_P^{-2}$}
\put(355,125){$\Lambda=0.2\ell_P^{-2}$}
\end{picture}
\caption{\footnotesize 
The left panel shows the expectation value for the bulk area squared and compares with the critical value of the action in the Euclidean regime. We observe that already the Regge expectation value deviates from the critical value of the classical action, with the difference increasing with larger scale factors.  The effective spin foam expectation value starts to deviate from the Regge expectation value, starting with quite small scale factors. A refinement of the spectrum for the bulk area increases the regime where Regge and (refined) spin foam expectation value coincide. 
\label{FigB2}}
\end{figure}

Fig.~\ref{FigB2} shows the spin foam expectation value for the bulk area squared and compares it to the Regge integral expectation value as well as to the classical\footnote{There is no classical solution, but we do have a saddle point along Euclidean data. We refer to the position of this saddle point as ``classical value".} value.

We again see that the effective spin foam value deviates from the Regge integral value starting with quite small boundary areas. Refining the spectrum (by a factor of 10), we can push the regime where we get agreement between spin foam and Regge result to larger boundary areas. We also find a deviation between the Regge result and the classical value, but they show the same qualitative   behaviour. 

For the second example we will choose $\Lambda=2\ell_P^{-2}$. We choose the same range of boundary values for the area as before. Due to having a cosmological constant value which is 10 times the size of the value in the first example, this range now covers 10 percent (and not only 1 percent) of the regime in which we approximate well continuum behaviour with the Regge path integral.

\begin{figure}[ht!]
\begin{picture}(500,150)
\put(0,2){ \includegraphics[scale=0.6]{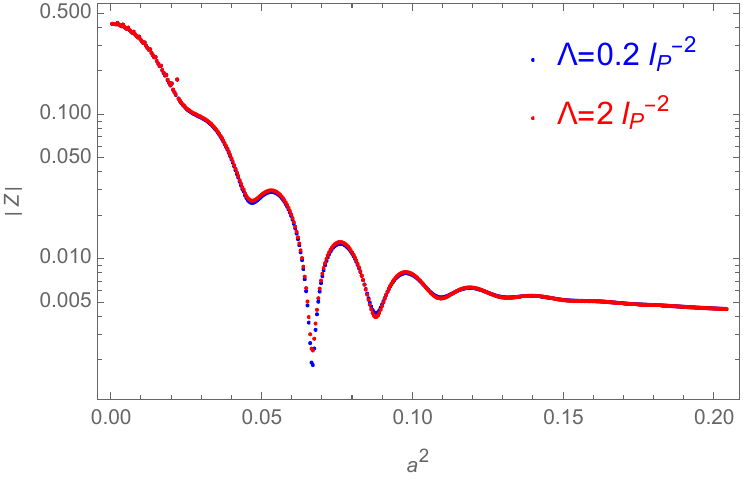} }
\put(250,2){ \includegraphics[scale=0.6]{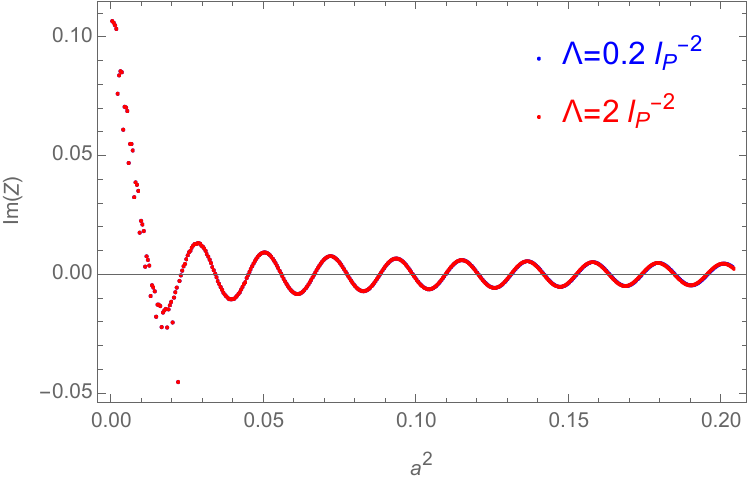} }
\end{picture}
\caption{\footnotesize  Here we compare the absolute value of the spin foam partition function  (left panel, logarithmic plot) and its  imaginary part (right panel) for the values  $\Lambda=0.2 \ell_P^{-2}$ and $\Lambda=2 \ell_P^{-2}$ of the cosmological constant. There are almost no differences visible in this plot.
\label{FigB3}}
\end{figure}

We noticed that the results for $\Lambda=2\ell_P^{-2}$ look very similar to the results for $\Lambda=0.2\ell_P^{-2}$ (for the same range of boundary areas). That is for these ranges of boundary areas the value of the cosmological constant has a  weak influence. We will therefore just show a comparison of the effective spin foam sum for these two cases in Fig.~\ref{FigB3}.

\begin{figure}[ht!]
\begin{picture}(500,150)
\put(0,2){ \includegraphics[scale=0.6]{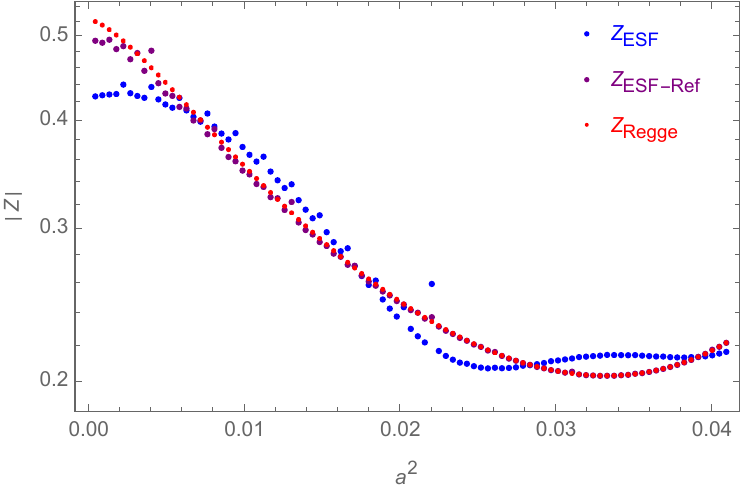} }
\put(250,2){ \includegraphics[scale=0.6]{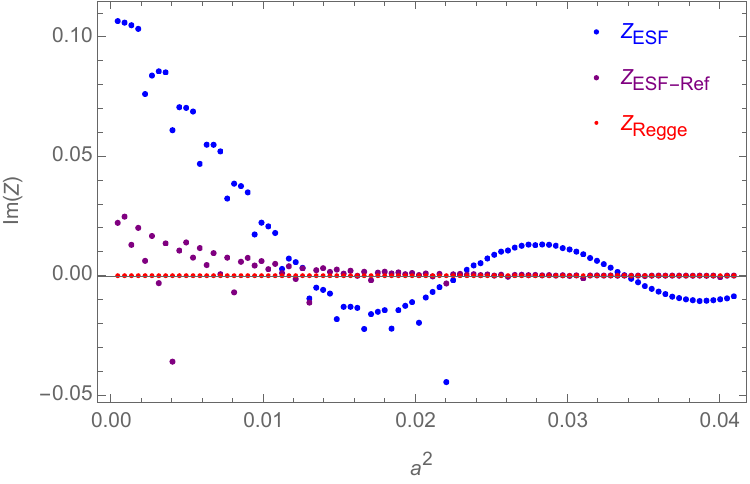} }
\put(95,125){$\Lambda=100\ell_P^{-2}$}
\put(355,125){$\Lambda=100\ell_P^{-2}$}
\end{picture}
\caption{\footnotesize  
Here we consider a very large cosmological constant $\Lambda=100\ell_P^{-2}$ and compare the absolute value (left panel, logarithmic plot) and imaginary part of the partition function for effective spin foams, effective spin foams with refined spectrum, and the Regge integral. We show the entire range of the boundary scale factor squared, for which the classical equations of motions for the discrete model emulate the continuum behaviour. This range scales with the inverse of $\Lambda$, and is therefore very small, including only 90 eigenvalues for the boundary area.  We observe that over  this small range the differences for the various versions of the partition function are small. 
\label{FigB4}}
\end{figure}

\begin{figure}[ht!]
\begin{picture}(500,150)
\put(0,2){ \includegraphics[scale=0.6]{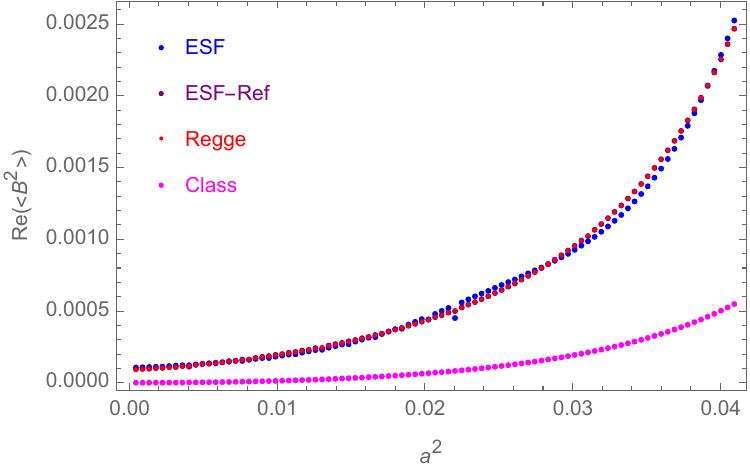} }
\put(250,2){ \includegraphics[scale=0.6]{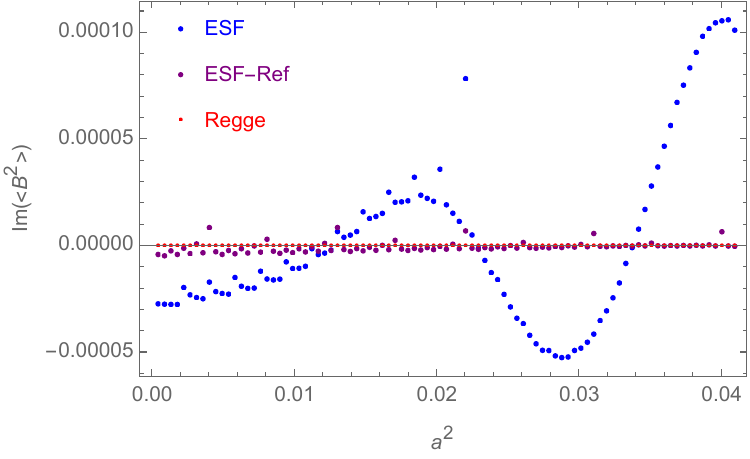} }
\put(95,120){$\Lambda=100\ell_P^{-2}$}
\put(355,120){$\Lambda=100\ell_P^{-2}$}
\end{picture}
\caption{\footnotesize  These plots shows the real and imaginary part of  the expectation value for the bulk area squared, and compare with the critical value of the classical action.  We see that (for the real part) the expectation value for the various versions of partition functions do almost agree, but all deviate from the classical value. \label{FigB5}}
\end{figure}

Increasing the cosmological constant further and further, we decrease the range of boundary values which lead to an Euclidean saddle point and show for the action evaluated on this saddle point the same qualitative behaviour as in the continuum \cite{DGS}. E.g. choosing a value of $\Lambda=100 \ell_P^2$, this range only extends to $A=90 (\gamma/n_t) \ell_P^2$. Here we find agreement between the effective spin foam and the Regge results, see Fig.~\ref{FigB4} and Fig.~\ref{FigB5}. But the reader might  want to keep in mind that we are looking at a range of boundary areas, which is five times smaller than in the previous examples, and that we found also agreement in this smaller range in the previous examples.

In summary, in the Euclidean regime we find rather large deviations between the effective spin foam sum and the Regge integral, in particular when comparing to the results in the Lorentzian regime. These deviations grow with the difference between the boundary areas. We still find that the effective spin foam amplitude describing the  (tunnelling) transition amplitude from a universe with vanishing scale factor to a universe with finite scale factor is suppressed. But, above a certain threshold value for the final scale factor  the probability for such transitions is increased in effective spin foams as compared to the Regge path integral.

\section{Discussion}\label{Discussion}

We discussed here a Lorentzian path integral for quantum gravity based on effective spin foams, and applied it
to a triangulation modelling homogeneous and isotropic cosmology.  Effective spin foams are much more amenable to numerical computations than previous spin foam models and have also no restrictions regarding the space-like or time-like nature of lower dimensional building blocks. This allowed us to perform the first cosmological spin foam calculations which integrate over a lapse like bulk variable. One main advantage of this feature is that it allows us a direct comparison of the results with the Regge path integral as well as with quantum mini-superspace cosmology. 

The evaluation of Lorentzian path integrals is very challenging due to the oscillating nature of the integrands. We have seen that the integrals and sums we consider are either very slowly converging (Fig.~\ref{FigAcc1}) or even diverging (Fig.~\ref{FigAcc2}). 

Integrals can be treated by deforming the contour into the complex plane, e.g. following the Picard-Lefschetz prescription \cite{PLTheory}. We do not have such a technique available for sums, and the theory of discrete harmonic functions does also not seem to help \cite{Unpublished}.  Non-linear transformations of sequences  can however accelerate the convergence of series very much and may also allow the extraction of (anti-) limits from diverging series \cite{WenigerReview}.  We discovered here that the  Shanks transforms (and Wynn's epsilon algorithm) are particularly effective in dealing with the series at hand, and that these can be applied to sums and integrals. It also allows the computation of expectation values, which involve divergent series. 

The main reason for the effectiveness of these techniques is the simple asymptotic behaviour of the Regge action. It seems in particular important that in the case of the sums the action is linear in the summation variable (see Footnote \ref{footnoteQ}). This is the case for the examples considered here but does hold also more generally for spin foams \cite{ReggeAsymptotics}.  We find it notable that  (asymptotically) equidistant area spectra (for $(3+1)$ dimensional gravity), as implemented in spin foams, does lead to an action linear in the summation variable, whereas (asymptotically) equidistant length spectra would lead to the problems outlined in Footnote \ref{footnoteQ}.  Keeping this lesson in mind, we hope that the Shanks transform and Wynn's epsilon algorithm can be applied to more general gravitational  partition functions as well as more general real time path integrals.  

We have applied the non-linear series transformation to compute the Regge path integral and the effective spin foam sums for two kinds of simplicial min-superspace models, namely the ball and shell model. 

To summarize the results we first note that the construction of the cosmological effective spin foam models ensures that a saddle point approximation reproduces the correct classical cosmological dynamics, within the regime where discretization artifacts can be neglected \cite{DGS}.\footnote{In contrast, the works \cite{SFC1,SFC2,SFC3}, which aimed to extract a cosmological dynamics from the EPRL model \cite{EPRLFK}, had to focus on this question, and did not include a (lapse-like) bulk variable.}

We therefore focused on the transition from the Regge path integral to the effective spin foam sum, that is the effects induced by implementing the discrete area spectrum.  We considered for the shell model boundary conditions that led to critical points along Lorentzian data. We found for this class of examples only small differences between the Regge path integral and the spin foam sum. Indeed the restriction to Lorentzian critical points means that we have to choose relatively large scale factors: in the continuum the threshold value is $a_\Lambda=\sqrt{3/\Lambda}$. We did find more significant deviations for very large cosmological constant $\Lambda=100\ell_P^2$.

The ball model models a transition from a vanishing scale factor to a finite scale factor and therefore implements the no-boundary proposal.\footnote{Here we find that the cases with Euclidean critical points lead to an exponentially suppressed amplitude (consistent with \cite{TurokEtAl,ADP}). Note that we decided to sum over the hinge irregular configurations with a choice of (complex) Regge action leading to exponentially suppressed amplitudes. We expect to find a different result if we implement the opposite choice. Finding an exponentially suppressed amplitude is compatible with the findings of \cite{HanLiu}, which apply a saddle point approximation for the one-simplex amplitude expressed as an integral over continuous parameters. On the other hand \cite{SFC4} does find a non-suppressed amplitude for a discretization given by one (not further subdivided) simplex. The reason are so-called vector geometries, which are thought to describe degenerate configurations \cite{BarrettAs}. It is still open whether such vector geometries should be allowed in the spin foam path integral or not \cite{EngleD}.} For the choice of the finite scale factor restricted to the regime in which the Regge discretization provides a good approximation to the continuum dynamics, which is governed by Euclidean critical points. In previous work \cite{ADP} we found that the Regge path integral, in this regime, reproduces very well the continuum mini-superspace results \cite{TurokEtAl}, despite the fact that the ball model describes only one time step.  In both cases one finds an exponential decay of the path integral. If one includes hinge irregular configurations into the integral, it does lead to a real partition function, as is also the case in the continuum.

The introduction of the discrete area spectrum (\ref{BulkSp}) does however, for the choices with larger finite scale factor,  lead to large deviations between the effective spin foam sum and the Regge path integral. The exponential decay (as a function of the outer scale factor) is very much weakened and the spin foam sum does acquire significant imaginary terms.

We can thus conclude the following: The examples with Lorentzian critical points react fairly insensitive to the introduction of a discrete area spectrum.  The examples with Euclidean critical points do lead to significant differences. We provided an heuristic explanation, based on the observation that although one can in principle apply a deformation of the `integration contour' also for sums, the discrete analytical continuation is much less under control than in the continuum \cite{Unpublished}.

The construction of our model did involve a number of simplifications, and lifting these simplifications might affect our conclusions. Future work will address these, e.g.:
\begin{itemize}
\item We provided heuristic arguments for the choice of a discrete area spectrum in (\ref{BulkSp}) in the symmetry reduced model. A similar issue appears for loop quantum cosmology \cite{LQCfromFull}. It would be desirable to have a more rigorous derivation of the spectrum in symmetry reduced models. One way to achieve this is to allow for more inhomogeneous bulk variables and to then integrate out the inhomogeneities and in this way to define an effective model for only symmetry reduced configurations. That is, to implement a coarse graining and renormalization process \cite{DittrichReview14,BahrFrusta,Review22}.
\item We only considered one time step. With the shell model we can easily incorporate more time steps, both in the Euclidean and Lorentzian regime. We found that for both the Lorentzian and Euclidean regime the deviations between Regge integral and spin foam sum got stronger with larger differences between initial and final scale factor. It will be interesting to see whether we arrive at the same conclusion if we incorporate more time steps, and how e.g. the threshold value beyond which we see strong deviations in the Euclidean regime, changes. 
\end{itemize}

A further advantage of introducing more time steps (and eventually perform the refinement limit) is that the refinement limit can also restore time reparametrization invariance. This has already been demonstrated for the classical dynamics in \cite{DGS}. We expect that this will also resolve ambiguities for the path integral measure, as has been illustrated in \cite{PathIntHO,MeasureRegge}. Here it will be in particular interesting to study the role of the irregular hinge configurations, e.g. whether it will be necessary to include these in order to achieve time reparametrization invariance.
With time reparametrization invariance restored, one can derive a Hamiltonian constraint and the effective spin foam partition function (when defined to include the sum over positive and negative lapse variable) will satisfy this constraint \cite{PathIntHO,Hoehn}.  (Note that the ball model does automatically satisfy a so called post constraint equation \cite{Hoehn}).

Other interesting extensions to the model presented here include the addition of matter and to study the possibility of singularity resolution. A simple dust action has already been considered in \cite{Collins1973,DGS}.  In a similar vein, it will be highly interesting to add more geometric variables which e.g. describe anisotropies \cite{DPToappear}. 

These efforts will lead to partition functions which involve the summation of several or even many variables. Let us note that e.g. the Shanks transformation can be generalized to examples with more than one summation variable. One can e.g. organize the multiple sums into one sequence \cite{Multidim}. A possible organization principle for path integrals is to integrate or sum first over all configurations whose actions evaluates to a value in a certain small range. This seems in particular promising, as the Regge gravitational action shows a simple asymptotic behaviour when all bulk areas or edge length are scaled large \cite{ReggeAsymptotics}.

\appendix

\section{Regge actions for the ball and shell model}\label{AppA}

Here we will provide a more detailed account of the Regge actions for the ball and shell model.

The action for the ball model has the following form
\ba\label{Action1App}
\pm\imath \,(8\pi G) S^{\pm}_{\rm Ball}&=& n_e \sqrt{\!\!{}_{{}_\pm} \mathbb{A}_{\rm blk}} \,\delta^{\pm}_{\rm blk} + n_t \sqrt{\!\!{}_{{}_\pm}\, \mathbb{A}_{\rm bdry}} \,\delta^{\pm}_{\rm bdry} -\Lambda n_\tau \sqrt{\!\!{}_{{}_\pm} \mathbb{V}_\sigma} \q ,
\ea
where the various geometric quantities are given by   
\ba\label{Ball2}
\sqrt{\!\!{}_{{}_\pm} \, \mathbb{A}_{\rm blk}} &=&\frac{\sqrt{s_l}}{4\sqrt{2}}\, \sqrt{\!\!{}_{{}_\pm} \,s_l+8 s_h} \q , \nn\\
\delta^\pm_{\rm blk}&=& 2\pi \mp 6\frac{n_\tau}{n_e}\imath \log_\mp
\left(
 \frac{-s_l+8s_h \mp\imath 8(s_l +8s_h)\sqrt{\!\!{}_{{}_\pm}\,\,  \frac{s_h}{s_l+8s_h} } }{s_l+24 s_h} 
 \right)         \q , \nn\\
\sqrt{\!\!{}_{{}_\pm} \, \mathbb{A}_{\rm bdry}} &=& \frac{\sqrt{3}}{4} s_l                    \q ,  \nn\\
\delta^\pm_{\rm bdry}&=& \pi \mp2 \imath \log_\mp
\left(
\frac{ 
\sqrt{s_l} \mp\imath 2\sqrt{6}\sqrt{ \!\!{}_{{}_\pm}\,\,  s_h}
}
{ 
\sqrt{\!\!{}_{{}_\pm} \,\,s_l+24 s_h}
}
\right)      \q , \nn\\
\sqrt{\!\!{}_{{}_\pm}\mathbb{V}_\sigma}  &=& \frac{\sqrt{2}}{48} s^{3/2}_l \sqrt{   \!\!{}_{{}_\pm}\,\,      s_h}        \q .
\ea   
The variable $s_l$ gives the squared length of the edges in the boundary of the 600-cell and $s_h$ gives the (signed) squared height of the four-simplices, into which the 600-cell is divided. We can replace $s_l$ and $s_h$ by the (signed) area squares $\mathbb{A}_{\rm bdry}$ and $\mathbb{A}_{\rm blk}$, using the relations in (\ref{Ball2}). We then obtain the action as a function of these area (squares).

We have for the shell model
\ba
\pm\imath \,(8\pi G) S^\pm_{\rm Shell}&=& n_e \sqrt{ \mathbb{A}_{\rm blk}} \,\delta^\pm_{\rm blk} + n_t  \left(\sqrt{\!\!{}_{{}_\pm} \mathbb{A}_{\rm bdry_1}} \, \,\delta^\pm_{\rm bdry_1}  + \sqrt{\!\!{}_{{}_\pm} \mathbb{A}_{\rm bdry_2}} \, \,\delta^\pm_{\rm bdry_2}  \right)   -\Lambda n_\tau \sqrt{\!\!{}_{{}_\pm}\mathbb{V}_{\text{4-frust}}}  \,\,\, ,
\ea
where we have now
\ba\label{Shell2}
\sqrt{\!\!{}_{{}_\pm} \, \mathbb{A}_{\rm blk}} &=&\frac{\sqrt{s_{l_1}}+ \sqrt{s_{l_2}  }}{4\sqrt{2}}\, \sqrt{\!\!{}_{{}_\pm} \,(\sqrt{s_{l_2}}-\sqrt{s_{l_1}})^2+8 s_h} \q , \nn\\
\delta^\pm_{\rm blk}&=& 2\pi \mp6
\frac{n_\tau}{n_e}\imath \log_\mp
\left(
 \frac{-(\sqrt{s_{l_2}}-\sqrt{s_{l_1}})^2+8s_h \mp\imath 8((\sqrt{s_{l_2}}-\sqrt{s_{l_1}})^2 +8s_h)\sqrt{\!\!{}_{{}_\pm}\,\,  \frac{s_h}{(\sqrt{s_{l_2}}-\sqrt{s_{l_1}})^2+8s_h} } }{(\sqrt{s_{l_2}}-\sqrt{s_{l_1}})^2+24 s_h} 
 \right)         \q , \nn\\
\sqrt{\!\!{}_{{}_\pm} \, \mathbb{A}_{\rm bdry_1}} &=& \frac{\sqrt{3}}{4} s_{l_1}                    \q ,  \q\q
\sqrt{\!\!{}_{{}_\pm} \, \mathbb{A}_{\rm bdry_2}} \,=\, \frac{\sqrt{3}}{4} s_{l_2}   \q ,  \nn\\
\delta^\pm_{\rm bdry_1}&=& \pi \mp2 \imath \log_\mp
\left(
\frac{ 
(\sqrt{s_{l_1}}-\sqrt{s_{l_2}}) \mp\imath 2\sqrt{6}\sqrt{ \!\!{}_{{}_\pm}\,\,  s_h}
}
{ 
\sqrt{\!\!{}_{{}_\pm} \,\,(\sqrt{s_{l_2}}-\sqrt{s_{l_1}})^2+24 s_h}
}
\right)       \, , \q
\delta^\pm_{\rm bdry_2}\,=\, \pi \mp2 \imath \log_\mp
\left(
\frac{ 
(\sqrt{s_{l_2}}-\sqrt{s_{l_1}}) \mp\imath 2\sqrt{6}\sqrt{ \!\!{}_{{}_\pm}\,\,  s_h}
}
{ 
\sqrt{\!\!{}_{{}_\pm} \,\,(\sqrt{s_{l_2}}-\sqrt{s_{l_1}})^2+24 s_h}
}
\right)       \q , \nn\\
\sqrt{\!\!{}_{{}_\pm}\mathbb{V}_{\text{4-frust}} } &=& \frac{\sqrt{2}}{48}  (\sqrt{s_{l_1}}+\sqrt{s_{l_2}}) (s_{l_2}+s_{l_1})    \sqrt{   \!\!{}_{{}_\pm}\,\,      s_h}        \q .
\ea
Here $s_{l_1}$ and $s_{l_2}$ give the squared edge lengths in the inner and outer boundary of the shell, respectively. The variable $s_h$ is the (signed) height square in the four-dimensional frusta, which build up the shell. We can again replace these variables by the (signed) squared areas $\mathbb{A}_{\rm bdry_1}$ and $\mathbb{A}_{\rm bdry_2}$ as well as $\mathbb{A}_{\rm blk}$.

\vspace{5mm}

\begin{acknowledgments}
 
JPA is supported by an NSERC grant awarded to BD.  Research at Perimeter Institute is supported in part by the Government of Canada through the Department of Innovation, Science and Economic Development Canada and by the Province of Ontario through the Ministry of Colleges and Universities.
\end{acknowledgments}

\providecommand{\href}[2]{#2}
\begingroup
\endgroup

\begin{thebibliography}{100}


\bibitem{deBoer} J.~de Boer, B.~Dittrich, A.~Eichhorn, S.~B.~Giddings, S.~Gielen, S.~Liberati, E.~R.~Livine, D.~Oriti, K.~Papadodimas and A.~D.~Pereira, \textit{et al.}
``Frontiers of Quantum Gravity: shared challenges, converging directions,''
[arXiv:2207.10618 [hep-th]].

\bibitem{CDT} J.~Ambjorn and R.~Loll,
``Nonperturbative Lorentzian quantum gravity, causality and topology change,''
Nucl. Phys. B \textbf{536} (1998), 407-434
[arXiv:hep-th/9805108 [hep-th]].
J.~Ambjorn, J.~Jurkiewicz and R.~Loll,
``A Nonperturbative Lorentzian path integral for gravity,''
Phys. Rev. Lett. \textbf{85} (2000), 924-927
[arXiv:hep-th/0002050 [hep-th]].
J.~Ambjorn, J.~Jurkiewicz and R.~Loll,
``Dynamically triangulating Lorentzian quantum gravity,''
Nucl. Phys. B \textbf{610} (2001), 347-382
[arXiv:hep-th/0105267 [hep-th]].

\bibitem{EPRLFK} 
  J.~Engle, E.~Livine, R.~Pereira and C.~Rovelli,
  ``LQG vertex with finite Immirzi parameter,''
  Nucl.\ Phys.\ B {\bf 799} (2008) 136
  [arXiv:0711.0146 [gr-qc]].
L.~Freidel and K.~Krasnov,
  ``A New Spin Foam Model for 4d Gravity,''
  Class.\ Quant.\ Grav.\  {\bf 25} (2008) 125018
  [arXiv:0708.1595 [gr-qc]].
  

\bibitem{PerezReview} A.~Perez,
  ``The Spin Foam Approach to Quantum Gravity,''
  Living Rev.\ Rel.\  {\bf 16} (2013) 3
  [arXiv:1205.2019].

\bibitem{TurokEtAl} J.~Feldbrugge, J.~L.~Lehners and N.~Turok,
``Lorentzian quantum cosmology,''
Phys.\ Rev.\ D \textbf{95} (2017), 103508,
[arXiv:1703.02076 [hep-th]].

\bibitem{EffSF3} S.~K.~Asante, B.~Dittrich and J.~Padua-Arg\"uelles,
``Effective spin foam models for Lorentzian quantum gravity,''
Class.\ Quant.\ Grav. in press (2021),
[arXiv:2104.00485 [gr-qc]].

\bibitem{Sato} Y.~Ito, D.~Kadoh and Y.~Sato,
``Tensor network approach to 2D Lorentzian quantum Regge calculus,''
Phys. Rev. D \textbf{106} (2022) no.10, 106004
[arXiv:2208.01571 [hep-th]].

\bibitem{ConfFac} G.~W.~Gibbons, S.~W.~Hawking and M.~J.~Perry,
``Path Integrals and the Indefiniteness of the Gravitational Action,''
Nucl. Phys. B \textbf{138} (1978), 141-150.

\bibitem{PLTheory} S.~Lefschetz, {\it Applications of algebraic topology, graphs and networks, the Picard-Lefschetz theory and Feynman integrals}, Applied Mathematical Sciences {\bf 16}, (Springer, Berlin, New York 1975)
V.~A.~Vassiliev, {\it Applied Picard-Lefschetz theory}, Mathematical Surveys and Monographs, {\bf 97} (AMS, Providence, R.I, 2002)

\bibitem{Witten1} E.~Witten,
``Analytic Continuation Of Chern-Simons Theory,''
AMS/IP Stud. Adv. Math. \textbf{50} (2011), 347-446
[arXiv:1001.2933 [hep-th]].
E.~Witten,
``A New Look At The Path Integral Of Quantum Mechanics,''
[arXiv:1009.6032 [hep-th]].

\bibitem{RelTime} Y.~Tanizaki and T.~Koike,
``Real-time Feynman path integral with Picard\textendash{}Lefschetz theory and its applications to quantum tunneling,''
Annals Phys. \textbf{351} (2014), 250-274
[arXiv:1406.2386 [math-ph]].

\bibitem{QCDReview} M.~Cristoforetti \textit{et al.} [AuroraScience],
``New approach to the sign problem in quantum field theories: High density QCD on a Lefschetz thimble,''
Phys. Rev. D \textbf{86} (2012), 074506
[arXiv:1205.3996 [hep-lat]].
L.~Bongiovanni,
``Numerical methods for the sign problem in Lattice Field Theory,''
[arXiv:1603.06458 [hep-lat]].
A.~Alexandru, G.~Basar, P.~F.~Bedaque and N.~C.~Warrington,
``Complex Paths Around The Sign Problem,''
[arXiv:2007.05436 [hep-lat]].
G.~Fujisawa, J.~Nishimura, K.~Sakai and A.~Yosprakob,
``Backpropagating Hybrid Monte Carlo algorithm for fast Lefschetz thimble calculations,''
[arXiv:2112.10519 [hep-lat]].

\bibitem{HanLefschetz} M.~Han, Z.~Huang, H.~Liu, D.~Qu and Y.~Wan,
``Spinfoam on a Lefschetz thimble: Markov chain Monte Carlo computation of a Lorentzian spinfoam propagator,''
Phys. Rev. D \textbf{103} (2021) no.8, 084026
[arXiv:2012.11515 [gr-qc]].

\bibitem{Ding2021} D.~Jia,
``Complex, Lorentzian, and Euclidean simplicial quantum gravity: numerical methods and physical prospects,''
[arXiv:2110.05953 [gr-qc]].

\bibitem{ADP} S.~K.~Asante, B.~Dittrich and J.~Padua-Arg\"uelles,
``Complex actions and causality violations: applications to Lorentzian quantum cosmology,''
Class. Quant. Grav. \textbf{40} (2023) no.10, 105005
[arXiv:2112.15387 [gr-qc]].


\bibitem{books}
C.~Rovelli,
``Quantum gravity,''
Univ. Pr., 2004,
doi:10.1017/CBO9780511755804 .
T.~Thiemann,
``Modern Canonical Quantum General Relativity,''
Cambridge University Press, 2007,
doi:10.1017/CBO9780511755682 .

\bibitem{LQCreviews}
M.~Bojowald,
``Loop quantum cosmology,''
Living Rev. Rel. \textbf{8} (2005), 11
[arXiv:gr-qc/0601085 [gr-qc]].
A.~Ashtekar and P.~Singh,
``Loop Quantum Cosmology: A Status Report,''
Class. Quant. Grav. \textbf{28} (2011), 213001
[arXiv:1108.0893 [gr-qc]].
K.~Banerjee, G.~Calcagni and M.~Martin-Benito,
``Introduction to loop quantum cosmology,''
SIGMA \textbf{8} (2012), 016
[arXiv:1109.6801 [gr-qc]].
I.~Agullo and P.~Singh,
``Loop Quantum Cosmology,''
[arXiv:1612.01236 [gr-qc]].

\bibitem{Bojowald}
M.~Bojowald,
``Critical evaluation of common claims in loop quantum cosmology,''
Universe \textbf{6} (2020) no.3, 36
[arXiv:2002.05703 [gr-qc]].
M.~Bojowald,
``Noncovariance of \textquotedblleft{}covariant polymerization\textquotedblright{} in models of loop quantum gravity,''
Phys. Rev. D \textbf{103} (2021) no.12, 126025
[arXiv:2102.11130 [gr-qc]].



\bibitem{Brunnemann:2007du}
J.~Brunnemann and C.~Fleischhack,
``On the configuration spaces of homogeneous loop quantum cosmology and loop quantum gravity,''
[arXiv:0709.1621 [math-ph]].
J.~Brunnemann and T.~A.~Koslowski,
``Symmetry Reduction of Loop Quantum Gravity,''
Class. Quant. Grav. \textbf{28} (2011), 245014
[arXiv:1012.0053 [gr-qc]].

\bibitem{SFC1} E.~Bianchi, C.~Rovelli and F.~Vidotto,
``Towards Spinfoam Cosmology,''
Phys. Rev. D \textbf{82} (2010), 084035
[arXiv:1003.3483 [gr-qc]].

\bibitem{Dona2022} P.~Dona and P.~Frisoni,
``How-to Compute EPRL Spin Foam Amplitudes,''
Universe \textbf{8} (2022) no.4, 208
[arXiv:2202.04360 [gr-qc]].

\bibitem{EffSF1} S.~K.~Asante, B.~Dittrich and H.~M.~Haggard,
``Effective Spin Foam Models for Four-Dimensional Quantum Gravity,''
Phys. Rev. Lett. \textbf{125} (2020) no.23, 231301
[arXiv:2004.07013 [gr-qc]].

\bibitem{EffSF2} S.~K.~Asante, B.~Dittrich and H.~M.~Haggard,
``Discrete gravity dynamics from effective spin foams,''
Class. Quant. Grav. \textbf{38} (2021) no.14, 145023
[arXiv:2011.14468 [gr-qc]].


\bibitem{SFC2} E.~Bianchi, T.~Krajewski, C.~Rovelli and F.~Vidotto,
``Cosmological constant in spinfoam cosmology,''
Phys. Rev. D \textbf{83} (2011), 104015
[arXiv:1101.4049 [gr-qc]].

\bibitem{SFC3} F.~Vidotto,
``Many-nodes/many-links spinfoam: the homogeneous and isotropic case,''
Class. Quant. Grav. \textbf{28} (2011), 245005
[arXiv:1107.2633 [gr-qc]].

\bibitem{SFC4} F.~Gozzini and F.~Vidotto,
``Primordial Fluctuations From Quantum Gravity,''
Front. Astron. Space Sci. \textbf{7} (2021), 629466
[arXiv:1906.02211 [gr-qc]].

\bibitem{SFC5} P.~Frisoni, F.~Gozzini and F.~Vidotto,
``Markov chain Monte Carlo methods for graph refinement in spinfoam cosmology,''
Class. Quant. Grav. \textbf{40} (2023) no.10, 105001
[arXiv:2207.02881 [gr-qc]].

\bibitem{Unpublished} B.~Dittrich, J.~Padua-Arg\"uelles, {\it unpublished notes}

\bibitem{WenigerReview} E.~J.~Weniger,
``Nonlinear sequence transformations for the acceleration of convergence and the summation of divergent series,"
Computer Physics Reports \textbf{10} (1989), 189,
	[arXiv:math/0306302 [math.NA]].




\bibitem{Conrady} F.~Conrady and J.~Hnybida,
``A spin foam model for general Lorentzian 4-geometries,''
Class. Quant. Grav. \textbf{27} (2010), 185011
[arXiv:1002.1959 [gr-qc]].

\bibitem{GielenOriti} S.~Gielen, D.~Oriti and L.~Sindoni,
``Cosmology from Group Field Theory Formalism for Quantum Gravity,''
Phys. Rev. Lett. \textbf{111} (2013) no.3, 031301
[arXiv:1303.3576 [gr-qc]].
S.~Gielen, D.~Oriti and L.~Sindoni,
``Homogeneous cosmologies as group field theory condensates,''
JHEP \textbf{06} (2014), 013
[arXiv:1311.1238 [gr-qc]].

\bibitem{DGS} B.~Dittrich, S.~Gielen and S.~Schander,
``Lorentzian quantum cosmology goes simplicial,''
Class. Quant. Grav. \textbf{39} (2022) no.3, 035012
[arXiv:2109.00875 [gr-qc]].

\bibitem{Regge} T.~Regge,
  ``General Relativity Without Coordinates,''
Nuovo Cim.  {\bf 19} (1961) 558.


\bibitem{Smolin} C.~Rovelli and L.~Smolin,
``Discreteness of area and volume in quantum gravity,''
Nucl. Phys. B \textbf{442} (1995), 593-622
[erratum: Nucl. Phys. B \textbf{456} (1995), 753-754]
[arXiv:gr-qc/9411005 [gr-qc]].

\bibitem{Ashtekar} A.~Ashtekar and J.~Lewandowski,
``Quantum theory of geometry. 1: Area operators,''
Class. Quant. Grav. \textbf{14} (1997), A55-A82
[arXiv:gr-qc/9602046 [gr-qc]].

\bibitem{Ryan1} B.~Dittrich and J.~P.~Ryan,
``Phase space descriptions for simplicial 4d geometries,''
Class. Quant. Grav. \textbf{28} (2011), 065006
[arXiv:0807.2806 [gr-qc]]. 
B.~Dittrich and J.~P.~Ryan,
``Simplicity in simplicial phase space,''
Phys. Rev. D \textbf{82} (2010), 064026
[arXiv:1006.4295 [gr-qc]].

\bibitem{AreaReggeLimit} B.~Dittrich,
``Modified Graviton Dynamics From Spin Foams: The Area Regge Action,''
[arXiv:2105.10808 [gr-qc]].
B.~Dittrich and A.~Kogios,
``From spin foams to area metric dynamics to gravitons,''
Class. Quant. Grav. \textbf{40} (2023) no.9, 095011
[arXiv:2203.02409 [gr-qc]].

\bibitem{JohannaArea} J.~N.~Borissova and B.~Dittrich,
``Towards effective actions for the continuum limit of spin foams,''
Class. Quant. Grav. \textbf{40} (2023) no.10, 105006
[arXiv:2207.03307 [gr-qc]].

\bibitem{JoseArea} B.~Dittrich and J.~Padua-Arg\"uelles,
``Twisted geometries are area-metric geometries,''
[arXiv:2302.11586 [gr-qc]].

\bibitem{AreaAngle} B.~Dittrich and S.~Speziale,
  ``Area-angle variables for general relativity,''
  New J.\ Phys.\  {\bf 10} (2008) 083006
  [arXiv:0802.0864 [gr-qc]].
  
  \bibitem{Asante:2022lnp}
S.~K.~Asante, J.~D.~Sim\~ao and S.~Steinhaus,
``Spin-foams as semiclassical vertices: Gluing constraints and a hybrid algorithm,''
Phys. Rev. D \textbf{107} (2023) no.4, 046002
[arXiv:2206.13540 [gr-qc]].

\bibitem{Han:2023cen}
M.~Han, H.~Liu and D.~Qu,
``Complex critical points in Lorentzian spinfoam quantum gravity: 4-simplex amplitude and effective dynamics on double-$\Delta_3$ complex,''
[arXiv:2301.02930 [gr-qc]].

\bibitem{Ryan3} B.~Dittrich and J.~P.~Ryan,
``On the role of the Barbero-Immirzi parameter in discrete quantum gravity,''
Class. Quant. Grav. \textbf{30} (2013), 095015
[arXiv:1209.4892 [gr-qc]].

\bibitem{HigherGauge} S.~K.~Asante, B.~Dittrich, F.~Girelli, A.~Riello and P.~Tsimiklis,
``Quantum geometry from higher gauge theory,''
Class. Quant. Grav. \textbf{37} (2020) no.20, 205001
[arXiv:1908.05970 [gr-qc]].

\bibitem{KBF} A.~Baratin and L.~Freidel,
``Hidden Quantum Gravity in 4-D Feynman diagrams: Emergence of spin foams,''
Class. Quant. Grav. \textbf{24} (2007), 2027-2060
[arXiv:hep-th/0611042 [hep-th]].
A.~Baratin and L.~Freidel,
``A 2-categorical state sum model,''
J. Math. Phys. \textbf{56} (2015) no.1, 011705
[arXiv:1409.3526 [math.QA]].

\bibitem{Simao} J.~D.~Sim\~ao and S.~Steinhaus,
``Asymptotic analysis of spin-foams with timelike faces in a new parametrization,''
Phys. Rev. D \textbf{104} (2021) no.12, 126001
[arXiv:2106.15635 [gr-qc]].

\bibitem{NumSFReview} P.~Dona, M.~Han and H.~Liu,
``Spinfoams and high performance computing,''
[arXiv:2212.14396 [gr-qc]].

\bibitem{AreaRegge} J.~W.~Barrett, M.~Rocek and R.~M.~Williams,
  ``A Note on area variables in Regge calculus,''
  Class.\ Quant.\ Grav.\  {\bf 16} (1999) 1373
  [gr-qc/9710056].

\bibitem{ADH} S.~K.~Asante, B.~Dittrich and H.~M.~Haggard,
``The Degrees of Freedom of Area Regge Calculus: Dynamics, Non-metricity, and Broken Diffeomorphisms,''
Class. Quant. Grav. \textbf{35} (2018) no.13, 135009
[arXiv:1802.09551 [gr-qc]].

\bibitem{BenLambda} B.~Bahr, G.~Rabuffo and S.~Steinhaus,
``Renormalization of symmetry restricted spin foam models with curvature in the asymptotic regime,''
Phys. Rev. D \textbf{98} (2018) no.10, 106026
[arXiv:1804.00023 [gr-qc]].

\bibitem{NewRegge} B.~Bahr and B.~Dittrich,
``Regge calculus from a new angle,''
New J. Phys. \textbf{12} (2010), 033010
[arXiv:0907.4325 [gr-qc]].

\bibitem{Lambda} W.~J.~Fairbairn and C.~Meusburger,
``Quantum deformation of two four-dimensional spin foam models,''
J. Math. Phys. \textbf{53} (2012), 022501
[arXiv:1012.4784 [gr-qc]].
M.~Dupuis and F.~Girelli,
``Observables in Loop Quantum Gravity with a cosmological constant,''
Phys. Rev. D \textbf{90} (2014) no.10, 104037
[arXiv:1311.6841 [gr-qc]].
H.~M.~Haggard, M.~Han, W.~Kami\'nski and A.~Riello,
``SL(2,C) Chern\textendash{}Simons theory, a non-planar graph operator, and 4D quantum gravity with a cosmological constant: Semiclassical geometry,''
Nucl. Phys. B \textbf{900} (2015), 1-79
[arXiv:1412.7546 [hep-th]].
B.~Dittrich,
``(3 + 1)-dimensional topological phases and self-dual quantum geometries encoded on Heegaard surfaces,''
JHEP \textbf{05} (2017), 123
[arXiv:1701.02037 [hep-th]].
M.~Han,
``Four-dimensional spinfoam quantum gravity with a cosmological constant: Finiteness and semiclassical limit,''
Phys. Rev. D \textbf{104} (2021) no.10, 104035
[arXiv:2109.00034 [gr-qc]].

\bibitem{Improve} B.~Bahr and B.~Dittrich,
``Improved and Perfect Actions in Discrete Gravity,''
Phys. Rev. D \textbf{80} (2009), 124030
[arXiv:0907.4323 [gr-qc]].
``Breaking and restoring of diffeomorphism symmetry in discrete gravity,''
AIP Conf. Proc. \textbf{1196} (2009) no.1, 10
[arXiv:0909.5688 [gr-qc]].

\bibitem{Review22} S.~K.~Asante, B.~Dittrich and S.~Steinhaus,
``Spin foams, Refinement limit and Renormalization,''
[arXiv:2211.09578 [gr-qc]].

\bibitem{DiffReview1} B.~Dittrich,
``Diffeomorphism symmetry in quantum gravity models,''
Adv. Sci. Lett. \textbf{2}, 151
[arXiv:0810.3594 [gr-qc]].

\bibitem{Broken} B.~Bahr and B.~Dittrich,
``(Broken) Gauge Symmetries and Constraints in Regge Calculus,''
Class. Quant. Grav. \textbf{26} (2009), 225011
[arXiv:0905.1670 [gr-qc]].

\bibitem{Hartle1985} J.~B.~Hartle,
``Simplicial minisuperspace I. General discussion,''
J.\ Math.\ Phys. \textbf{26} (1985), 804-814;
J.~B.~Hartle,
``Simplicial minisuperspace. II. Some classical solutions on simple triangulations,''
J.\ Math.\ Phys. \textbf{27} (1986), 287-295;
J.~B.~Hartle,
``Simplicial minisuperspace. III. Integration contours in a five-simplex model,''
J.\ Math.\ Phys. \textbf{30} (1989), 452-460.

\bibitem{Collins1973} P.~A.~Collins and R.~M.~Williams, ``Dynamics of the Friedmann Universe Using Regge Calculus'', Phys.\ Rev.\ D {\bf 7} (1973), 965-971.
R.~G.~Liu and R.~M.~Williams, ``Regge calculus models of the closed vacuum $\Lambda$--FLRW universe,'' Phys.\ Rev.\ D \textbf{93} (2016), 024032


\bibitem{Dittrich:2024awu}
B.~Dittrich, T.~Jacobson and J.~Padua-Arg\"uelles,
``De Sitter horizon entropy from a simplicial Lorentzian path integral,''
[arXiv:2403.02119 [gr-qc]].

\bibitem{HartleEtAl}
J.~Diaz Dorronsoro, J.~J.~Halliwell, J.~B.~Hartle, T.~Hertog and O.~Janssen,
``Real no-boundary wave function in Lorentzian quantum cosmology,''
Phys. Rev. D \textbf{96} (2017) no.4, 043505
[arXiv:1705.05340 [gr-qc]].


\bibitem{BahrFrusta} B.~Bahr, S.~Kloser and G.~Rabuffo,
``Towards a Cosmological subsector of Spin Foam Quantum Gravity,''
Phys. Rev. D \textbf{96} (2017) no.8, 086009
[arXiv:1704.03691 [gr-qc]].

\bibitem{LollLC} S.~Jordan and R.~Loll,
``Causal Dynamical Triangulations without Preferred Foliation,''
Phys. Lett. B \textbf{724} (2013), 155-159
[arXiv:1305.4582 [hep-th]]. 
S.~Jordan and R.~Loll,
``De Sitter Universe from Causal Dynamical Triangulations without Preferred Foliation,''
Phys. Rev. D \textbf{88} (2013), 044055
[arXiv:1307.5469 [hep-th]].

\bibitem{Sorkin1974} R.~Sorkin,
``Time Evolution Problem in Regge Calculus,''
Phys. Rev. D \textbf{12} (1975), 385-396
[erratum: Phys. Rev. D \textbf{23} (1981), 565-565]

\bibitem{Sorkin2019} R.~D.~Sorkin,
``Lorentzian angles and trigonometry including lightlike vectors,''
[arXiv:1908.10022 [gr-qc]].

\bibitem{HartleSorkin} J.~B.~Hartle and R.~Sorkin,
``Boundary Terms in the Action for the Regge Calculus,''
Gen. Rel. Grav. \textbf{13} (1981), 541-549.

\bibitem{BojowaldLivRev} M.~Bojowald,
``Loop quantum cosmology,''
Living Rev. Rel. \textbf{11} (2008), 4

\bibitem{Mubar} A.~Ashtekar, T.~Pawlowski and P.~Singh,
``Quantum Nature of the Big Bang: Improved dynamics,''
Phys. Rev. D \textbf{74} (2006), 084003
[arXiv:gr-qc/0607039 [gr-qc]].

\bibitem{Schmidt} J.~R.~Schmidt, ``On the numerical solution of linear simultaneous equations by an iterative
method," Philos. Mag. \textbf{32} (1941), 369.

\bibitem{Shanks} D.~Shanks, ``Non-linear transformations of divergent and slowly convergent sequences," J. Math.
and Phys. \textbf{34} (1955), 1.

\bibitem{Aitken} A.~Aitken, “On Bernoulli’s Numerical Solution of Algebraic Equations,” Proceedings of the Royal Society
of Edinburgh \textbf{46} (1927) 289-305.

\bibitem{Wynn} P.~Wynn, ``On a device for computing the $e_m(S_n)$ transformation," Math. Tables Aids Comput.
\textbf{10} (1956), 91.

\bibitem{Wynn2} P.~Wynn, ``A note on programming repeated applications of the $\epsilon$-algorithm," R.F.T.I. - Chiffres
\textbf{8} (1965), 23 .

\bibitem{Dona2} P.~Don\`a and P.~Frisoni,
``Summing bulk quantum numbers with Monte~Carlo in spin foam theories,''
Phys. Rev. D \textbf{107} (2023) no.10, 106008
[arXiv:2302.00072 [gr-qc]].

\bibitem{HartleHawking} J.~B.~Hartle and S.~W.~Hawking,
``Wave Function of the Universe,''
Phys. Rev. D \textbf{28} (1983), 2960-2975

\bibitem{DiscreteHarm} R.~J.~Duffin, ``Basic properties of discrete analytic functions," Duke Mathematical Journal, Duke Math. J. \textbf{23} (1956), 335.

\bibitem{LQCfromFull} J.~Brunnemann and T.~A.~Koslowski,
``Symmetry Reduction of Loop Quantum Gravity,''
Class. Quant. Grav. \textbf{28} (2011), 245014
[arXiv:1012.0053 [gr-qc]].
C.~Beetle, J.~S.~Engle, M.~E.~Hogan and P.~Mendon\c{c}a,
``Diffeomorphism invariant cosmological sector in loop quantum gravity,''
Class. Quant. Grav. \textbf{34} (2017) no.22, 225009
[arXiv:1706.02424 [gr-qc]].

\bibitem{ReggeAsymptotics} J.~Borissova, B.~Dittrich, D.~Qu, M.~Schiffer, ``Asymptotics of Lorentzian quantum Regge calculus", {\it to appear}

\bibitem{HanLiu} M.~Han and H.~Liu,
``Analytic Continuation of Spin foam Models,''
[arXiv:2104.06902 [gr-qc]].

\bibitem{BarrettAs} J.~W.~Barrett, R.~J.~Dowdall, W.~J.~Fairbairn, F.~Hellmann and R.~Pereira,
``Lorentzian spin foam amplitudes: Graphical calculus and asymptotics,''
Class. Quant. Grav. \textbf{27} (2010), 165009
[arXiv:0907.2440 [gr-qc]].

\bibitem{EngleD} J.~Engle,
``A spin-foam vertex amplitude with the correct semiclassical limit,''
Phys. Lett. B \textbf{724} (2013), 333-337
[arXiv:1201.2187 [gr-qc]].

\bibitem{DittrichReview14} B.~Dittrich,
``From the discrete to the continuous: Towards a cylindrically consistent dynamics,''
New J. Phys. \textbf{14} (2012), 123004
[arXiv:1205.6127 [gr-qc]].
B.~Dittrich,
``The continuum limit of loop quantum gravity - a framework for solving the theory,''
[arXiv:1409.1450 [gr-qc]].

\bibitem{PathIntHO} B.~Bahr, B.~Dittrich and S.~Steinhaus,
``Perfect discretization of reparametrization invariant path integrals,''
Phys. Rev. D \textbf{83} (2011), 105026
[arXiv:1101.4775 [gr-qc]].

\bibitem{MeasureRegge} B.~Dittrich and S.~Steinhaus,
``Path integral measure and triangulation independence in discrete gravity,''
Phys. Rev. D \textbf{85} (2012), 044032
[arXiv:1110.6866 [gr-qc]].
J.~N.~Borissova and B.~Dittrich,
``Lorentzian quantum gravity via Pachner moves: one-loop evaluation,''
[arXiv:2303.07367 [hep-th]].

\bibitem{Hoehn} B.~Dittrich and P.~A.~Hohn,
``From covariant to canonical formulations of discrete gravity,''
Class. Quant. Grav. \textbf{27} (2010), 155001
[arXiv:0912.1817 [gr-qc]].
B.~Dittrich and P.~A.~Hoehn,
``Constraint analysis for variational discrete systems,''
J. Math. Phys. \textbf{54} (2013), 093505
[arXiv:1303.4294 [math-ph]].

\bibitem{DPToappear} B.~Dittrich and J.~Padua-Arg\"uelles, ``Bianchi I quantum cosmology from effective spin foams", {\it to appear}

\bibitem{Multidim} A.~C.~Genz, ``A nonlinear method for the acceleration of the convergence of multidimensional squences", Journal of Computational and Applied Mathematics \textbf{3} (1977), 181.


\end{thebibliography}
\end{document}